\documentclass[preprint,12pt]{elsarticle}

\usepackage{amssymb}
\usepackage{amsmath} 
\usepackage{color} 
\usepackage{mathtools} 

\usepackage[colorlinks,
            linkcolor=olive,
            anchorcolor=green,
            citecolor=teal
            ]{hyperref}

\journal{International Journal of Multiphase flow}
\hypersetup{draft}
\begin{document}

\begin{frontmatter}



\title{Inertial particle velocity and distribution in vertical turbulent channel flow: a numerical and experimental comparison}


\author{Guiquan~Wang$^{1}$, Kee~Onn~Fong$^{2,3}$, Filippo~Coletti$^{2,3}$, Jesse~Capecelatro$^{4}$}
\author{David~Richter\corref{cor1} $^{1}$ }

\ead{David.Richter.26@nd.edu}

\address{1. Department of Civil and Environmental Engineering and Earth Sciences, University of Notre Dame, Notre Dame, Indiana 46556, USA
\\ 
2. Department of Aerospace Engineering and Mechanics, University of Minnesota, Minneapolis, MN 55455, USA
\\
3. St. Anthony Falls Laboratory, University of Minnesota, Minneapolis, MN 55414, USA
\\
4. Department of Mechanical Engineering, University of Michigan, Ann Arbor, MI 48109, USA }

\begin{abstract}

This study is concerned with the statistics of vertical turbulent channel flow laden with inertial particles for two different volume concentrations ($\Phi_{V} = 3 \times 10^{-6}$ and $\Phi_{V} = 5 \times 10^{-5}$) at a Stokes number of $St^{+} = 58.6$ based on viscous units. Two independent direct numerical simulation models utilizing the point-particle approach are compared to recent experimental measurements, where all relevant nondimensional parameters are directly matched. While both numerical models are built on the same general approach, details of the implementations are different, particularly regarding how two-way coupling is represented. At low volume loading, both numerical models are in general agreement with the experimental measurements, with certain exceptions near the walls for the wall-normal particle velocity fluctuations. At high loading, these discrepancies are increased, and it is found that particle clustering is overpredicted in the simulations as compared to the experimental observations. Potential reasons for the discrepancies are discussed. As this study is among the first to perform one-to-one comparisons of particle-laden flow statistics between numerical models and experiments, it suggests that continued efforts are required to reconcile differences between the observed behavior and numerical predictions. 

\end{abstract}

\begin{keyword}
Inertial particles \sep wall turbulence \sep simulations \sep experiment


\end{keyword}

\end{frontmatter}


\section{Introduction}
\label{1.Introduction}


Over the last several decades, a large number of experimental studies have been dedicated to understanding fluid-particle interactions in turbulent channel flows \citep{balachandar2010turbulent}. In the near-wall region, \cite{kaftori1995particle1, kaftori1995particle2} observed that the behavior of particles is correlated with near-wall coherent structures in the dilute, near-neutral buoyancy limit ($\rho_p/\rho_f=1.05, ~St^+=0.065-18$ where $St^{+}$ is the particle Stokes number based on wall units). In the core region of the channel, \cite{fessler1994preferential} found that particles form clusters of length scale O(10 $\eta$), where $\eta$ is the Kolmogorov length scale for moderate inertia particles ($St^+=27-150$) with mass fractions ranging from $\Phi_m=0.03-1.0$. \cite{kulick1994particle} investigated the turbulence modification by high inertia particles ($St^+=292-2030$) by comparing each over a range of mass loading up to $\Phi_m=0.8$. Furthermore, \cite{benson2005effects} studied the effect of mass loading and wall roughness for high inertia particles ($St^+=2630$), which was further numerically investigated by \citet{capecelatro2015mass} and \citet{vreman2015turbulence}. Recently, \cite{FongJFM2019} studied in detail the particle spatial distribution both close to the wall and in the centerline of a vertical channel, along with series of particle statistics at relatively low Reynolds numbers and multiple mass fractions. With moderate Stokes number ($St^+=64-130$), they found a significant difference (particle distribution and particle fluctuation velocity) between mass loading $\Phi_m=6\times 10^{-3}$ and $\Phi_m=0.1$.

Despite the aforementioned experimental progress, measurement of the carrier-phase velocity field near the particles remains a major problem in studying turbulence modulation due to the presence of particles. At the same time, numerical approaches have become a powerful tool to help fill the entire parameter space and understand statistics and mechanisms that are difficult to observe. In turbulent dispersed multiphase flow computations, the Lagrangian point-particle approach coupled with direct numerical simulation (DNS) has been able to successfully capture certain phenomena such as particles' preferential accumulation and modulation of turbulence \citep{poelma2006particle, balachandar2010turbulent}, but challenges still persist in achieving quantitative prediction.  

According to the standard paradigm, when $\Phi_v$ is small (in the range of $\Phi_v \leq 10^{-6}$ ), the particles have a negligible effect on the turbulence (i.e. one-way coupling). Here, particles are transported by turbulent motions, and efforts have been aimed at describing this dispersion process. \cite{marchioli2002mechanisms}, for instance, associated particle re-entrainment mechanisms with the strongly coherent ejections and sweeps. Meanwhile, \cite{narayanan2003mechanisms} found that particles ($St^+=5,~15$) preferentially accumulate in the so-called low-speed streaks and tend to deposit on the wall in an open channel flow. In order to compare different numerical predictions, the low order statistics of both particle and carrier phase is benchmarked by \cite{marchioli2008statistics} for one-way coupled turbulent channel flows.

When $\Phi_v$ is moderate (in the regime of $10^{-6} \leq \Phi_v \leq 10^{-3}$), the particles can have a considerable effect on turbulence through momentum exchange (i.e. two-way coupling), especially when particles accumulate in certain regions of the flow. In comparison with one-way coupling, similar particle deposition behaviour has been observed over a wide particle parameter space ($St^+=0.055-0.889$ in \cite{pan1995numerical}; $St^+=1-100$ in \cite{sardina2012wall}; $St^+=1-100$ in \cite{nilsen2013voronoi}; $St^+=8.5-714$ in \cite{richter2013momentum} and $St^+=4.44-444$ in \cite{wang2019modulation}). However, \cite{li2001numerical} and \cite{nasr2009dns} showed that two-way coupling weakened the preferential distribution of particles compared with one-way coupling in the channel flow. In addition to turbulence modulation, low-inertia particles ($St^+=O(1)$) induce a destabilization effect on transition from laminar to turbulent flow, whereas large-Stokes-number particles ($St^+>O(10)$) actually stabilize the turbulence \cite{klinkenberg2011numerical, wang2019modulation}; this is sometimes accompanied by observed drag reduction in numerical models \citep{li2001numerical, dritselis2008numerical}.

With further increase of particle loading, collision between particles takes place and modifies both particle and fluid statistics (i.e. four-way coupling). The particle/particle collision weakens the preferential distribution of particles \citep{li2001numerical, nasr2009dns}, and reduces maximum near-wall concentrations \citep{kuerten2015effect}. The particle/wall collisions also affect the particle-induced turbulence modulation \citep{vreman2015turbulence}. In addition, particles at high mass loading tend to decrease the thickness of the boundary layer and increase the skin friction \citep{vreman2009two}, and act as the primary source of turbulence generation \citep{capecelatro2016strongly, capecelatro2018transition}. 

Overall, this broad range of numerical investigations has relied heavily on the use of the point-force approximation, which is widely applied for systems with large numbers of small, heavy particles; see \cite{li2001numerical}; \cite{klinkenberg2013numerical}; \cite{zhao2013interphasial}; \cite{gualtieri2013clustering}; \cite{lee2015modification}; \cite{vreman2015turbulence}; \cite{wang_richter_2019}. While experimental and numerical efforts have made significant progress in understanding the complex problem of particle-laden turbulent channel flow, there remains a continued lack of comparison and validation between consistent numerical and experimental observations \citep{eaton2009two}. This is especially for moderate inertia ($St^+=O(10)$) particles with volume concentrations in the regime of two-way coupling and four-way coupling, since most comparisons have been done for high Stokes number(e.g. \citet{benson2005effects}). Under these circumstances, preferential concentration and turbophoresis are at play, particle/particle and particle/wall collision might take place, and particles may modify fluid momentum. In this context, we leverage the recent experimental data of \citet{FongJFM2019} and perform a statistical comparison between DNS simulations from independent numerical codes (considering both two- and four-way coupling), particularly focusing on particle statistics and clustering behaviour. We aim at investigating $(i)$ particle/particle and particle/wall collisions; $(ii)$ the difference between the numerical predictions of a traditional point-force method with a more advanced volume-filtering method; and $(iii)$ the discrepancies between numerical simulations and experimental results in a Reynolds number, Stokes number, and mass fraction regime which can be achieved using DNS.\\

\section{Simulation Method}
\label{2.Simulation Method and Validation}

This study is based on comparing two different numerical models (\citet{richter2013momentum} and \citet{capecelatro2013euler}) to the experiments of \citet{FongJFM2019}. Here we describe the two DNS-based models.

\subsection{Point particle method}

In this section, the numerical method of \citet{richter2013momentum} is introduced and compared with existing simulations in the literature, including \cite{zhao2013interphasial}, \cite{capecelatro2015mass}, \cite{vreman2015turbulence}, as well as experimental measurements from \cite{kulick1994particle}, \cite{paris2001turbulence} and \cite{benson2005effects}. The purpose here is to first compare against existing data before performing our more detailed validation below.

\begin{figure*}[!ht]
\centering
\includegraphics[width=16 cm]{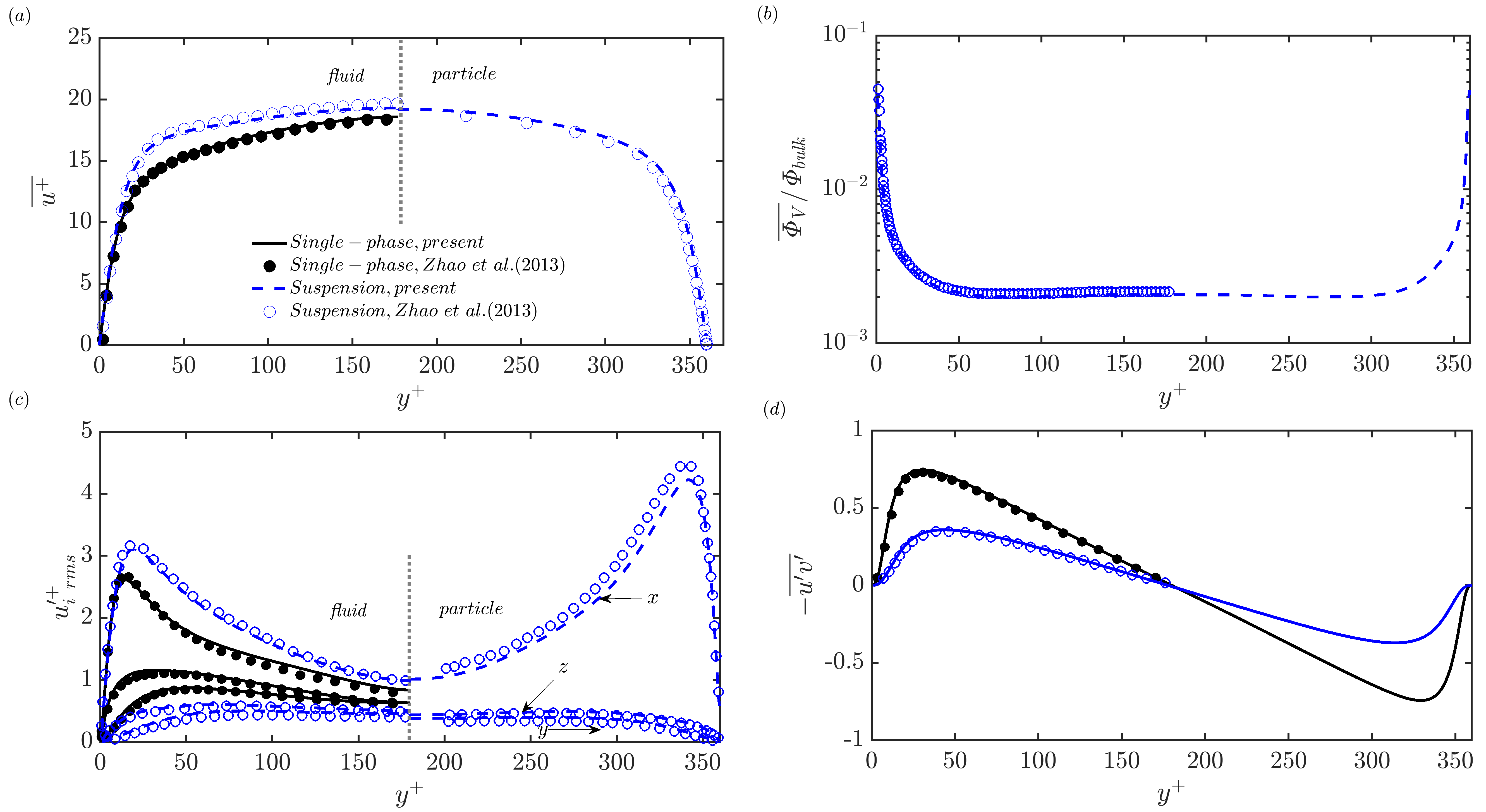} 
\caption{In comparison with \cite{zhao2013interphasial} pressure-driven channel flow at $Re_\tau=180$ laden with particles of $St^+=30$. $(a)$ Mean velocity profile in wall normal direction: left half is the fluid phase and right half is the particle phase; $(b)$ Particle volume fraction normalized by the bulk; $(c)$ RMS velocity fluctuation in wall normal direction: left half is the fluid phase and right half is the particle phase; $(d)$ Reynolds shear stress of the fluid.}
\label{fig:VS_Zhao}
\end{figure*}
Direct numerical simulations of single-phase flows are performed for an incompressible Newtonian fluid, and this model has been utilized previously in other studies \citep{richter2013momentum,richter2014modification}. A pseudospectral method is employed in the periodic directions (streamwise $x$ and spanwise $z$), and second-order finite differences are used for spatial discretization in wall-normal, $y$ direction. The solution is advanced in time by a third-order Runge-Kutta scheme. Incompressibility is achieved via the solution of a pressure Poisson equation. The fluid velocity and pressure fields are a solution of the continuity and momentum balance equations in Eqs. (\ref{eq:method_conti}) and (\ref{eq:method_momen}), respectively:

\begin{equation}\label{eq:method_conti}
\frac{\partial u_j}{\partial x_j}=0,
\end{equation} 

\begin{equation}\label{eq:method_momen}
\frac{\partial u_i}{\partial t} + u_j\frac{\partial u_i}{\partial x_j}=-\frac{1}{\rho_f} \frac{\partial p}{\partial x_i}+\nu \frac{\partial u_i}{\partial x_j \partial x_j} + \frac{1}{\rho_f} F_i +\delta_{i1} g.
\end{equation} 

Here $u_{i}$ is the fluid velocity, $p$ is the pressure, $F_i$ is the particle feedback force to the carrier phase computed by projecting the particle force to the nearest Eulerian grid points, $g$ is the acceleration of gravity, $\nu$ is the fluid kinematic viscosity, and $\rho_{f}$ is the fluid density. 

Particle trajectories and particle-laden flow dynamics are based on the point-force approximation where the particle-to-fluid density ratio $r \equiv \rho_p/\rho_f \gg 1$ and the particle size is smaller than the smallest viscous dissipation scales of the turbulence. As a consequence of this and the low volume concentrations (a maximum bulk volume fraction of $\overline{\Phi_{V}}$ less than $1 \times 10^{-3}$), only the Schiller-Naumann \citep{schiller1933ber} hydrodynamic drag force is considered. The velocity of particle $n$ is governed by Eq. (\ref{eq:method_drag_f}) and particle trajectories are then obtained from numerical integration of the equation of motion in Eq.  (\ref{eq:method_motion}):

\begin{equation}\label{eq:method_drag_f}
\frac{d u^n_{p,i}}{dt}={f^n_i} + f_i^c + \delta_{i1} g,
\end{equation} 

\begin{equation}\label{eq:method_motion}
\frac{dx^n_i}{dt}=u^n_{p,i},
\end{equation} 

where the drag is given by

\begin{equation}\label{eq:method_drag_f_2}
{f^n_i} =\frac{1}{\tau_p}[1+0.15(Re^n _p)^{0.687}] (u^n_{f,i}-u^n_{p,i}).
\end{equation} 

\noindent Here, $\tau _p = \rho _p {d_p} ^2 / 18\mu$ is the Stokes relaxation time of the particle, and the particle Reynolds number $Re^n_p=\mid u^n_{f,i}-u^n_{p,i}\mid d^n_p / \nu $ is based on the magnitude of the particle slip velocity $(u^n_{f,i}-u^n_{p,i})$ and particle diameter $d_{p}^{n}$. In this work, the average $Re^n_p$ is less than $1.0$, which is far smaller than the suggested maximum $Re_{p} \approx 800$ for the Stokes drag correction in Eq. (\ref{eq:method_drag_f}) \citep{schiller1933ber}. As a result of the low $Re_{p}$, the correction to the Stokes drag is minimal in this study. Other terms in the particle momentum equation \citep[see][]{maxey1983equation} are neglected since they remain small compared with drag when the density ratio $r \gg 1$. In all simulations, particles are initially distributed at random locations throughout the channel.  \\

In the two-way coupling configuration, particle-particle collisions are not taken into consideration, and we exert a purely elastic collision between particles and the upper/lower walls. This purely elastic wall collision is commonly used in gas-solid turbulence \citep{li2001numerical, sardina2012wall, zhao2013interphasial}, however we have tested the restitution coefficient $|u^n_{p,init}/u^n_{p,final}|$ between $0.5$ and $1$ and do not observe significant changes to particle distributions or two-way coupling, consistent with \citet{li2001numerical}. To demonstrate that our implementation provides results that are consistent with other similar numerical models, we provide a comparison to the two-way coupled simulations of \cite{zhao2013interphasial} in Fig. \ref{fig:VS_Zhao}. In this test, gravity is not considered, pressure-driven channel turbulence is simulated at a moderate friction Reynolds number of $Re_\tau=180$ (based on the friction velocity $u_{\tau}$), and the particle Stokes number normalized by viscous units is $St^+=30$. The particle concentration profile, mean velocity, RMS fluctuation velocity, and Reynolds shear stress are shown in Fig. \ref{fig:VS_Zhao} as a function of wall-normal distance. Here we are essentially confirming that the particle-force method is correctly implemented in the code and that it provides nearly identical results to other similar formulations in the two-way coupled regime.\\


\begin{figure*}[!ht]
\centering
\includegraphics[width=16 cm]{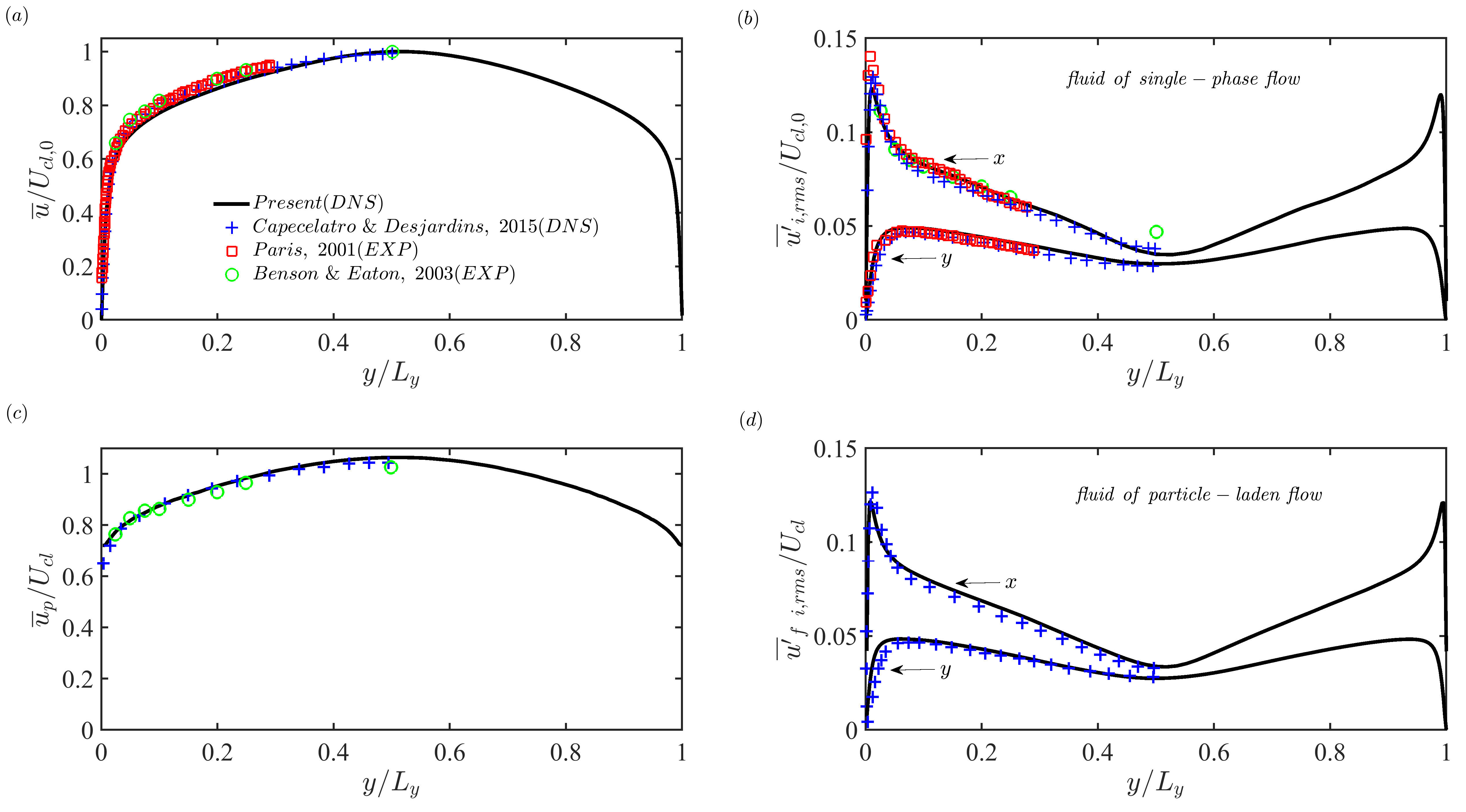} 
\caption{$(a)$ Mean fluid velocity profile in wall normal direction in single-phase flow; $(b)$ fluid RMS velocity fluctuation in wall normal direction in single-phase flow; $(c)$ Mean particle velocity profile in wall normal direction; $(d)$ RMS fluid velocity fluctuation in wall normal direction in particle-laden flow. All figures are normalized by the fluid centerline velocity.}
\label{fig:VS_Jesse_1}
\end{figure*}

\begin{figure*}[!ht]
\centering
\includegraphics[width=16 cm]{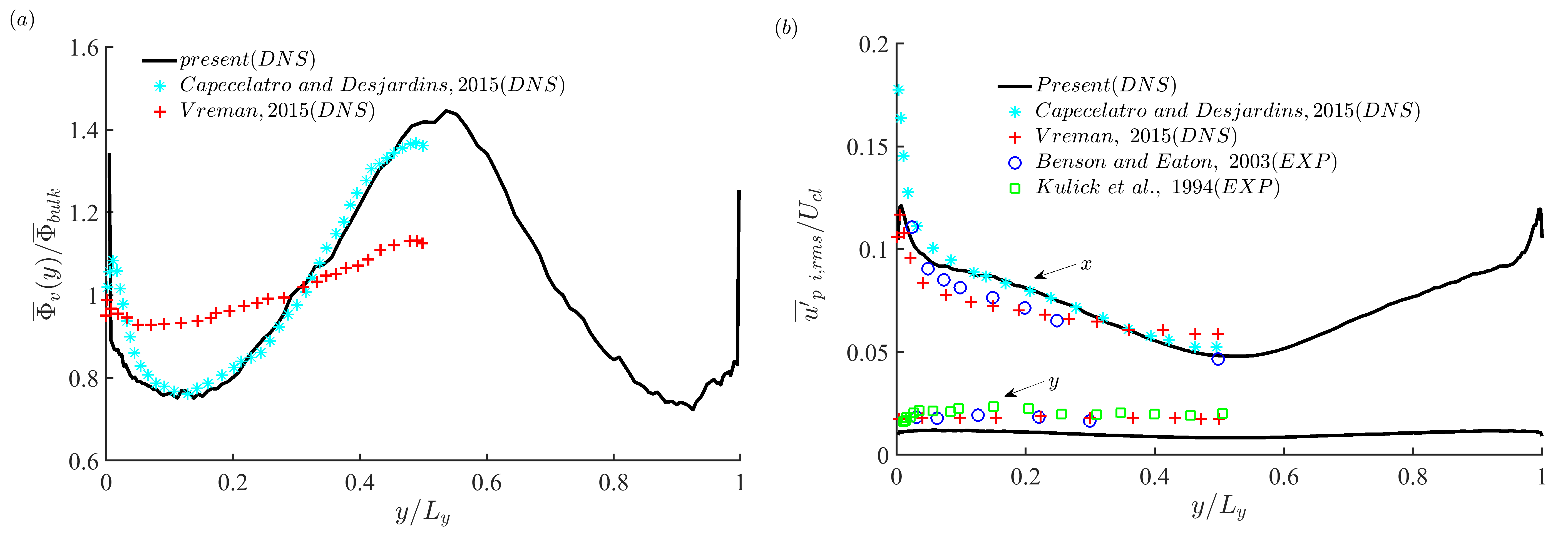} 
\caption{$(a)$ Particle concentration in wall normal direction, the mass loading ($\Phi_v=7.2 \times 10^{-5},~\Phi_m=0.15$) is same in present study and \cite{capecelatro2015mass}, while higher mass loading ($\Phi_v=9.1 \times 10^{-5},~\Phi_m=0.8$) is used in \cite{vreman2015turbulence}; $(b)$ RMS fluid velocity fluctuation in wall normal direction normalized by the fluid centerline velocity.}
\label{fig:VS_Jesse_2}
\end{figure*}


From the two-way coupling formulation described above, four-way coupling can be included as well, where particle/particle and particle/wall collisions are modeled according to a spring-dashpot system. The collision force $f_i^c$ in Eq. (\ref{eq:method_drag_f}) is computed by coupling the DNS code to the open source, DEM-based LIGGGHTS package for discrete element methods, applied initially by \cite{kloss2012models}. A Hertz-Mindlin contact model is used in the normal and tangential directions to the vector connecting particle centers. In the present study, we set the parameters in the collision model as follows: Young's modulus ($5 \times 10^{-5}$), Poisson's ratio ($0.45$), friction coefficient ($0.1$) and restitution coefficient ($0.9$). Further details on the numerical implementation and validation of the collision model can be found in \cite{kloss2012models}. \\

\subsection{Voume-averaged particle method}

This work also utilizes the model of \cite{capecelatro2013euler}, whose notable difference with the model of \citet{richter2013momentum} described above is that volume-averaging is used to apply the two-way coupling forces back to the Eulerian mesh; the model of \citet{richter2013momentum} uses the traditional particle-in-cell method, projecting only to the nearest nodes.

The volume-averaged Navier--Stokes equations employed in the model of \cite{capecelatro2013euler} are given by
\begin{equation}\label{fluid_continuity}
\frac{\partial\alpha}{\partial t}+\frac{\partial}{\partial x_i}(\alpha u_i) = 0
\end{equation}

and

\begin{equation}\label{fluid_mom}
\frac{\partial\alpha u_i}{\partial t}+\frac{\partial}{\partial x_i} \left(\alpha u_i u_j\right) =-\frac{1}{\rho_f}\frac{\partial p}{\partial x_i}+\frac{\partial \sigma_{ij}}{\partial x_j}-\frac{\rho_p}{\rho_f}\alpha_p F_i+\alpha\delta_{i1}g,
\end{equation}

where $\alpha$ is the fluid-phase volume fraction and $\alpha_p=1-\alpha$. The fluid-phase viscous-stress tensor is defined as

\begin{equation}\label{sigma}
\mathbf{\sigma}_{ij}=\left(\nu+\nu^\star\right)\left[\frac{\partial u_i}{\partial x_j}+\frac{\partial u_j}{\partial x_i}-\frac{2}{3}\frac{\partial u_k}{\partial x_k}\delta_{ij}\right]
\end{equation}

where $\nu^\star$ is an effective viscosity that accounts for enhanced dissipation due to unresolved fluid-velocity fluctuations generated at the particle scale \citep{gibilaro2007apparent}. Unlike in the point-particle description, in this model two-way coupling accounts for both the resolved stresses (pressure and viscous stress) and unresolved fluid stresses (i.e., drag). Thus, the momentum exchange term felt by particle $n$ is

\begin{equation}\label{part_source}
{f^n_i} =\frac{1}{\tau_p}[1+0.15(Re^n _p)^{0.687}] (u^n_{f,i}-u^n_{p,i})-\frac{1}{\rho_p}\frac{\partial p^n}{\partial x_i}+\frac{1}{\rho_p}\frac{\partial \sigma_{ij}^n}{\partial x_j}.
\end{equation} 

The momentum exchange term is projected to the grid via

\begin{equation}\label{alpha_A2}
\alpha_p F_i = \sum_{n=1}^{N_p}f^{n}_i\mathcal{G}(|x_i-x_i^{n}|)V_p,
\end{equation}

where $V_p=\pi d_p^3/6$ is the particle volume and $\mathcal{G}$ is a Gaussian kernel with characteristic size $\delta_f=8d_p$. This expression replaces the discontinuous Lagrangian data with an Eulerian field that is a smooth function of the spatial coordinate $x_i$. Similarly, the fluid volume fraction is computed as

\begin{equation}\label{alpha}
\alpha =1- \sum_{n=1}^{N_p}\mathcal{G}(|x_i-x_i^{n}|)V_p.
\end{equation}

To further test the numerical formulations against existing data, we perform a comparison with the simulations of \cite{capecelatro2015mass}  (focusing only on the dilute regime), which uses the model of \citet{capecelatro2013euler}. At the same time, we also compare to experimental results from \cite{kulick1994particle}, \cite{paris2001turbulence}, and \cite{benson2005effects} under similar conditions, despite the results of \citet{kulick1994particle} being subject to unconstrained roughness effects. In this test, turbulent, vertical channel flow is simulated at a high Reynolds number ($Re_\tau=630$) with high particle Stokes number ($St^+=2030$) and a mass loading of $\Phi_m=0.15$. Figures \ref{fig:VS_Jesse_1}(a,b) show the mean fluid velocity and RMS fluctuation velocity in single-phase flow; both numerical models agree well with the measurements from \cite{paris2001turbulence} and \cite{benson2005effects}. In particle-laden flow, the mean particle velocity compares well between numerical models and the measurement from \cite{benson2005effects}, and the computed RMS fluid fluctuation velocity agrees well between numerical models, as shown in figures \ref{fig:VS_Jesse_1}(c,d), respectively.

The particle concentration and particle-phase RMS fluctuation velocity are shown in Figs \ref{fig:VS_Jesse_2}(a) and (b), respectively. In Fig. \ref{fig:VS_Jesse_2}(a), the concentration profiles between the point particle model and the volume-averaged model of \cite{capecelatro2015mass} agree well with each other. In Fig \ref{fig:VS_Jesse_2}(b), \cite{capecelatro2015mass} compute a higher RMS particle streamwise velocity compared to the point particle model. In addition, the simulations exhibit a slightly lower RMS wall-normal velocity in the whole channel as compared to the experimental observations of \cite{kulick1994particle} and \cite{benson2005effects}. As a reference, simulation results from the model of \cite{vreman2015turbulence} are also included in Fig \ref{fig:VS_Jesse_2}, although the simulations of \cite{vreman2015turbulence} are at a much higher mass fraction ($\Phi_m=0.7$). The concentration profile is flatter in this case, while the RMS particle velocities are of similar magnitude.
\\

\section{Flow and particle parameters}
\label{3.parameters}

We now turn our attention towards a more detailed validation, which is based on the recent experiments of \cite{FongJFM2019}. The flow configuration of interest is pressure-driven, downwards-oriented channel flow (see \cite{FongJFM2019}). In the simulations, periodic boundary conditions are applied to both phases in the streamwise ($x$) and spanwise ($z$) directions. In \cite{FongJFM2019} two flow Reynolds numbers are used, and we focus on the $Re_{bulk} = 6020$ case, where $Re_{bulk} = 2h U_{bulk} / \nu$ is based on the bulk velocity $U_{bulk}$ and channel height $2h$. This approximately corresponds to $Re_{\tau}=227$ based on the friction velocity and $h$. For the experimental density ratio $\rho_{p}/\rho_{f} = 2083$ and diameter $d_{p} = 4.7 \times 10^{-5} \operatorname{m}$, this corresponds to a Stokes number of $St^{+} = 58.6$ based on viscous units and $St_{\eta} = 6.7$ based on the Kolmogorov scale at the centerline. Two experimental volume loadings are simulated: $\Phi_V = 3 \times 10^{-6}$ (``low'') and $\Phi_V = 5 \times 10^{-5}$ (``high'').\\


The fluid-phase flow parameters are provided in table \ref{tab:Table_1}. Throughout, the notation ``\_R'' and ``\_C'' refer to the models of \citet{richter2013momentum} and \cite{capecelatro2013euler}, respectively, which were described in section \ref{2.Simulation Method and Validation} . Both simulations were designed to accurately predict the unladen experiments and match the key nondimensional parameters. The relevant particle parameters are listed in table \ref{tab:Table_2}.  The particle diameter is smaller than the Kolmogorov scale $\eta$ ($d_p/\eta \sim 0.25$), and the conventional understanding is that the point-particle method should yield accurate predictions in this regime. \\

\begin{table} [!ht]
\centering
\def\arraystretch{1.0}
\caption{Fluid phase parameters for two DNS codes} 

\begin{tabular}{ccc}
\hline
                            	  & $Unladen\_R$                 & $Unladen\_C$                 \\ \hline
$h(m)$	        			      & \multicolumn{2}{c}{$15\times 10^{-3}$}                    
\\
$U_{cl}(m/s)$				      & \multicolumn{2}{c}{4.40}                    
\\
$U_{bulk}(m/s)$				      & \multicolumn{2}{c}{3.0}                    
\\
$Re_{bulk}$                       & \multicolumn{2}{c}{6020}                                  
\\
$Re_\tau$                    	  & \multicolumn{2}{c}{227}                                  
\\
$u_\tau$				          & \multicolumn{2}{c}{0.227}                                                     
\\
$\delta_\nu$				      & \multicolumn{2}{c}{$6.6 \times 10^{-5}$}
\\
$\tau_\nu$		         		  & \multicolumn{2}{c}{$2.9 \times 10^{-4}$}
\\
$\eta$		         		      & \multicolumn{2}{c}{$2.0 \times 10^{-4}(centerline)$}
\\
$\tau_\eta$		         		  & \multicolumn{2}{c}{$2.6 \times 10^{-3}(centerline)$}
\\
$L^+_x \times L^+_y \times L^+_z$ & \multicolumn{2}{c}{$4276 \times 454 \times 712$}    \\
$N_x \times N_y \times N_z$ 	  & $512 \times 128 \times 128$ & $656 \times 110 \times 110$ \\
$\Delta x^+$, $\Delta z^+$        & 8.35, 5.57                  & 6.52, 6.48                  \\ \hline
\end{tabular}

\label{tab:Table_1}
\end{table}

\begin{table} [!ht]
\centering
\def\arraystretch{1.0}
\caption{Particle parameters. $N_{p}$ refers to the total number of particles and $V_{s} \equiv \tau_p g$ refers to the terminal settling velocity in still fluid aligned in the streamwise direction, where $g$ is the gravitational acceleration.} 

\begin{tabular}{ccc}
\hline
                  & low concentration  & high concentration    \\ \hline
$\rho_p/\rho_f$ & \multicolumn{2}{c}{2083}                   \\
$d_p(m)$             & \multicolumn{2}{c}{$4.7 \times 10^{-5}$}   \\
$d_p^+$           & \multicolumn{2}{c}{$0.71$}                 \\
$d_p/\eta$        & \multicolumn{2}{c}{$0.25(centerline)$}                 \\
$\tau_p$          & \multicolumn{2}{c}{$0.017$}                 \\
$V_s(m/s)$          & \multicolumn{2}{c}{$0.167$}                 \\
$St^+$            & \multicolumn{2}{c}{$58.6$}                 \\
$St_\eta$         & \multicolumn{2}{c}{$6.7(centerline)$}                  \\
$\Phi_v$          & $3 \times 10^{-6}$ & $5 \times 10^{-5}$ \\
$N_p$             & $2.2 \times 10^{4}$ & $3.67 \times 10^{5}$ \\
$\Phi_m$          & $6.25 \times 10^{-3}$ & $0.1$ \\ \hline
\end{tabular}
\label{tab:Table_2}
\end{table}

An overview of the simulations conducted in this study is provided in table \ref{tab:Table_3}. For the model described above, we compare two- and four-way coupling against the experimental data and the four-way coupled model of \citet{capecelatro2013euler}. A smaller time step is used when four-way coupling is included so that collisions can be resolved. The total simulation time is more than $21h/u_\tau$ and the time for collecting statistics is at least $18h/u_\tau$.

\begin{table} [!ht]
\centering
\def\arraystretch{1.0}
\caption{Simulations conducted in this study} 

\begin{tabular}{cccc}
\hline
             & $\Phi_v$ & method                  & $\Delta t^+$ \\ \hline
$Unladen\_R$    & --       & --                      & 0.2          \\
$low\_2\_R$  & low      & two-way, point-force    & 0.2          \\
$high\_2\_R$ & high     & two-way, point-force    & 0.2          \\
$low\_4\_R$  & low      & four-way, point-force   & 0.1          \\
$high\_4\_R$ & high     & four-way, point-force   & 0.1          \\
$Unladen\_C$    & --       & --                      & 0.13         \\
$low\_4\_C$  & low      & four-way, volume-filtering & 0.033        \\
$high\_4\_C$ & high     & four-way, volume-filtering & 0.03         \\ \hline
\end{tabular}

\label{tab:Table_3}
\end{table}

\section{Preliminary comparisons}

\subsection{Unladen flow}
\begin{figure*}[!ht]
\centering
\includegraphics[width=16 cm]{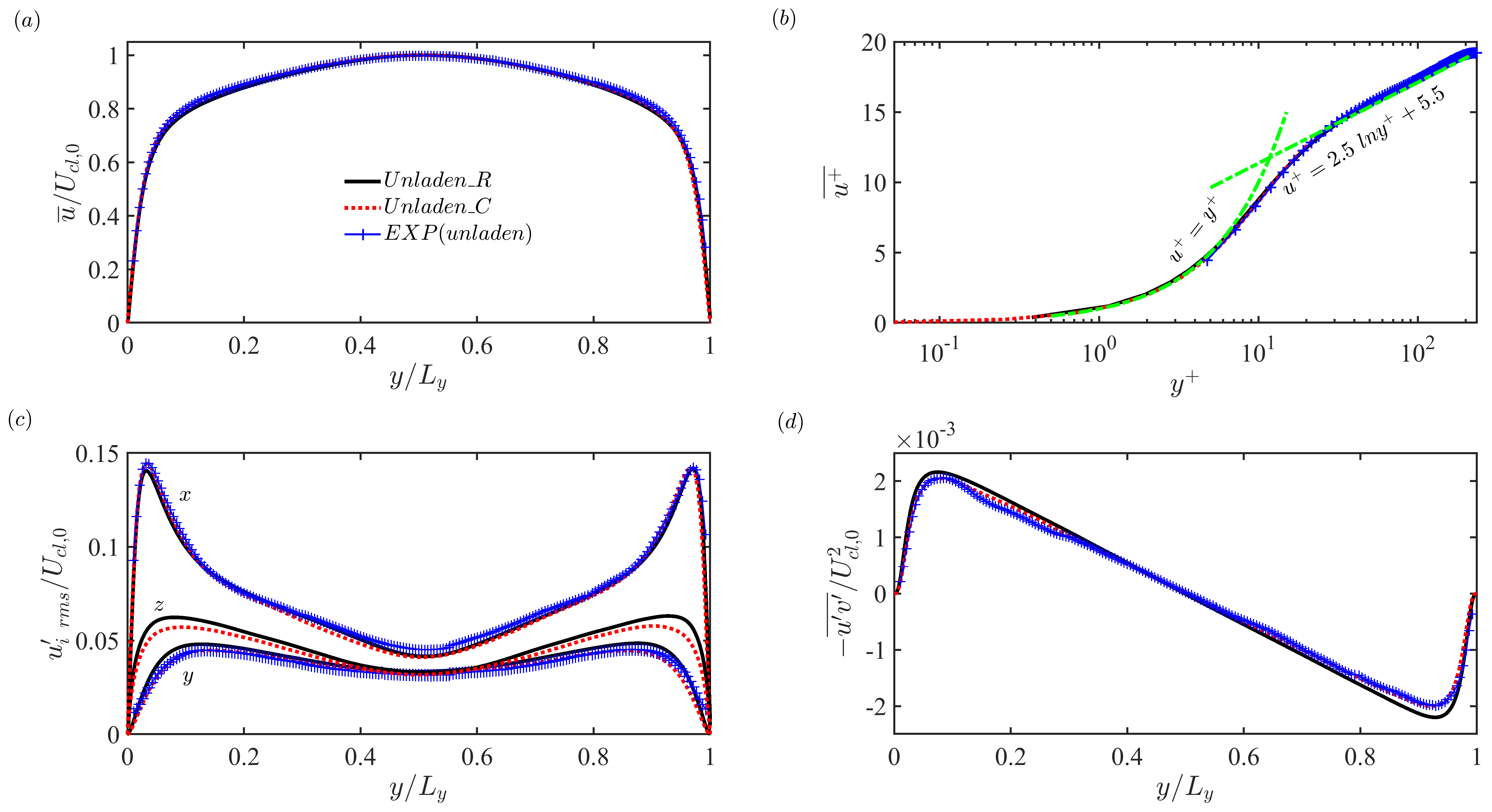} 
\caption{Unladen velocity statistics comparison between Unladen\_R, Unladen\_C and the experimental observations of \cite{FongJFM2019} as a function of wall-normal direction: mean fluid velocity profile in outer units $(a)$ and viscous units $(b)$; $(c)$ RMS fluid velocity fluctuation in three directions; $(d)$ Reynolds shear stress. All figures are normalized by the fluid centerline velocity $U_{cl}$.}
\label{fig:Single-phase}
\end{figure*}

Comparisons between both numerical models and the measurement produce nearly identical mean velocity profiles, shown in figs. \ref{fig:Single-phase}(a,b). In addition, the turbulent intensity profiles and Reynolds shear stress are shown in figs. \ref{fig:Single-phase}(c,d), respectively. In the near-wall region, $u'_{rms}$ agrees well whereas $v'_{rms}$, $w'_{rms}$ and $-\overline{u'v'}$ are slightly higher in $Unladen\_R$ than for $Unladen\_C$ and the measurement. In the center region, all components compare well between both numerical models and the experimental observations. The above comparisons indicate that both DNS codes have successfully captured the large-scale coherent structures in current configuration.\\


\subsection{Time evolution}
\label{3.2 Time evolution}

Before calculating time-averaged statistics, we first investigate the time required to achieve a statistically steady state of the particles as they transition from the uniform initial condition. Two particle timescales are involved: the first is the Stokes timescale $\tau_{p}$, which indicates how quickly particles can adjust to the local fluid velocity. The second is that associated with mean drift to an equilibrium mean profile, which is typically longer than $\tau_p$. From one-way coupled simulations, \cite{marchioli2008statistics} found that this time scale is longer for lower inertia particles, e.g., particles with $St^+=5$ spend three times as long as particles with a higher $St^+=25$ establishing a stationary concentration distribution. The time scale for particles moving to the equilibrium locations influences the length of the development section when performing an experiment. For example \cite{kulick1994particle} (respectively \cite{FongJFM2019}) designed the length scale of the development section to be five times (respectively twenty times) longer than the particle relaxation time scale multiplied by the centerline fluid velocity with Stokes number of $St^+=2030$ (respectively $St^+=58.6$).

\begin{figure}[!ht]
\centering
\includegraphics[width=8 cm]{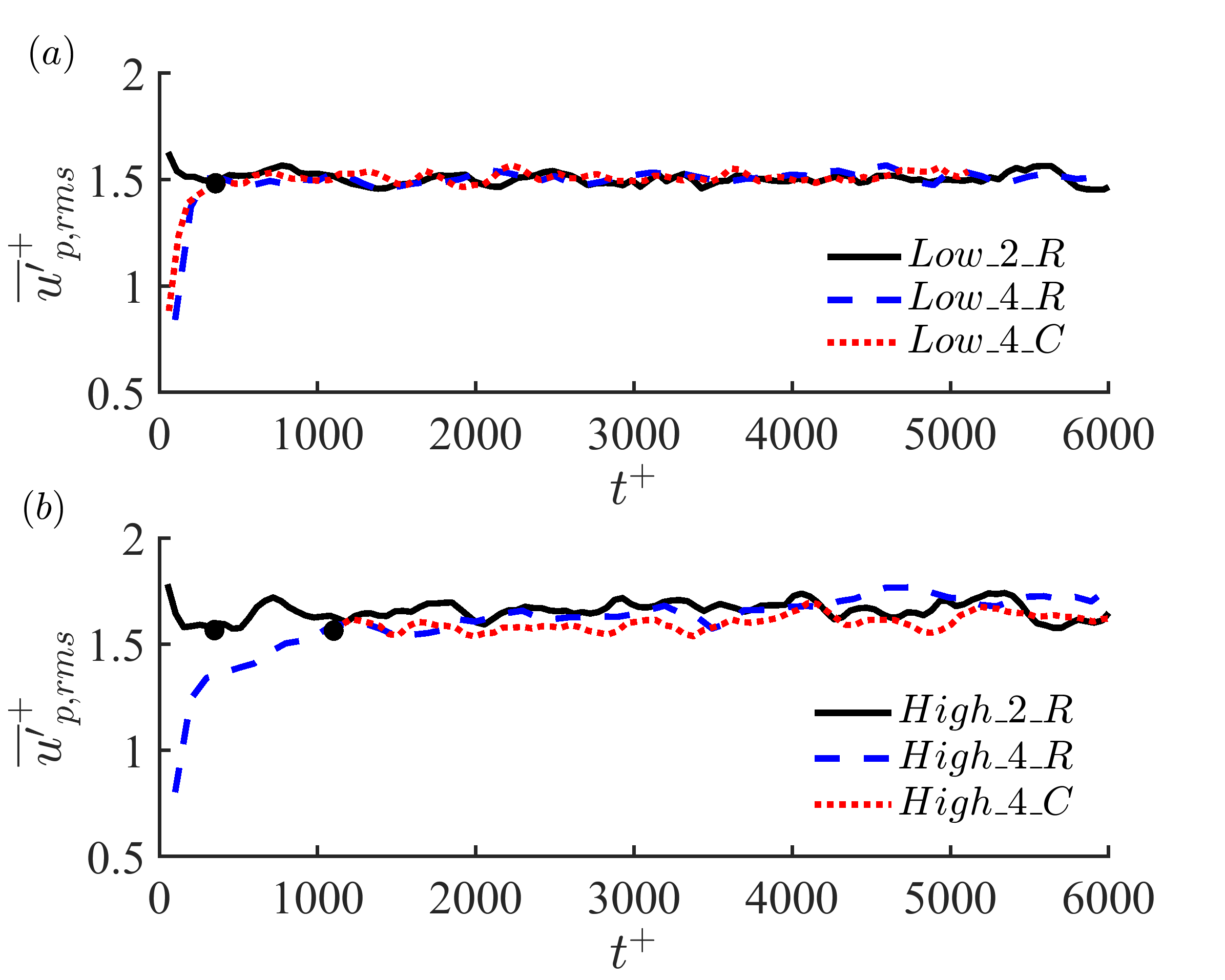} 
\caption{The domain averaged RMS particle streamwise fluctuation velocity as function of time in $(a)$ low concentration and $(b)$ high concentration.}
\label{fig:urms_evolution}
\end{figure}

Fig. \ref{fig:urms_evolution} shows the time evolution of the domain-averaged RMS particle fluctuation velocity, scaled by the viscous time and velocity scales, respectively. We can see that particle/particle collisions have very little effect on the time to stationarity ($t^+ \sim 400$) at low mass loading, as shown in Fig. \ref{fig:urms_evolution}(a). Comparing Figs. \ref{fig:urms_evolution}(a) and (b), the time required to achieve statistical stationarity at high mass loading is increased by collisions between particles ($t^+\sim 1200$). For the dimensional values used in the simulations and as a reference for future experiments, these results would suggest that the length of a development section should exceed five times of $\tau_p U_{cl}$ at low mass loading, and twenty times of $\tau_p U_{cl}$ at high mass loading, assuming that particles are randomly distributed initially in a fully-developed turbulent flow field.
\begin{figure}[!ht]
\centering
\includegraphics[width=7.5 cm]{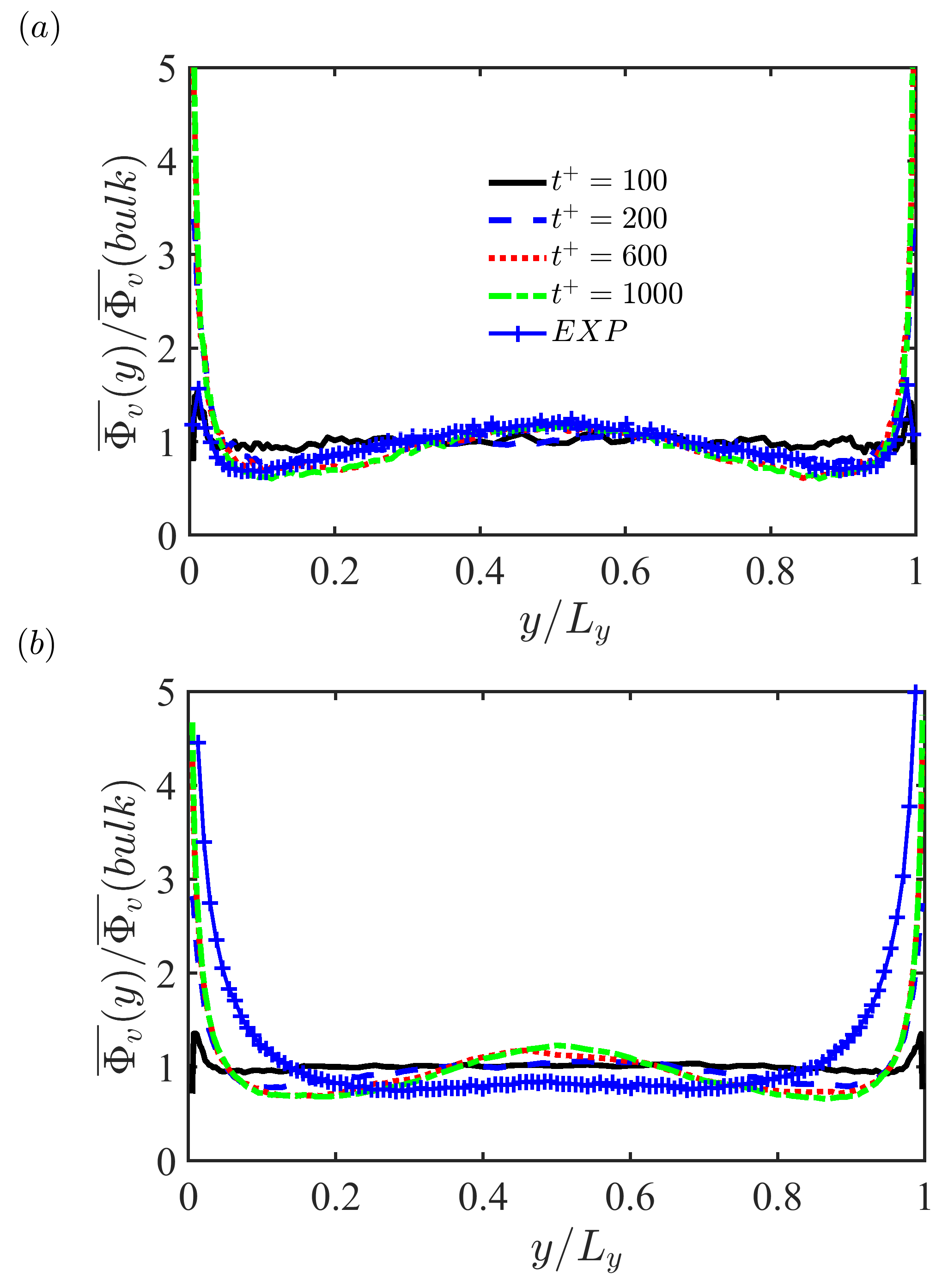} 
\caption{Particle concentration profiles in wall-normal direction at $t^+=100,~200,~600,~1000$. $(a)$ low concentration; $(b)$ high concentration. The measurements of \cite{FongJFM2019} are shown.}
\label{fig:Phi_evolution}
\end{figure}

To further emphasize this point, the time evolutions of the concentration profiles for cases $high\_2\_R$ and $high\_4\_R$ are shown in Figs. \ref{fig:Phi_evolution}(a) and (b), corresponding to the low and high concentrations, and these are compared to experimental measurements. From the simulations, we can see that for both concentrations, particles experience turbophoretic drift, where local maxima are found in concentration near the walls and at the centerline. This drift is enhanced due to the alignment of gravity in the mean flow direction, see for example \cite{capecelatro2015mass}. The time required to reach a stationary concentration profile is similar to that seen in figure \ref{fig:urms_evolution} for both mass loadings. As compared to the experiments, there is a strong agreement between simulations and observations at low mass loading, except at the wall where simulations overpredict the concentration by nearly a factor of five. At high mass loading, however, the simulations indicate a nearly identical evolution in time and corresponding steady-state concentration profile, while the experiments exhibit a marked change in cross-channel particle distribution at high mass loading. This will be further discussed in section \ref{4.1 Concentration}.

\subsection{Particle accumulation at the walls}
\label{3.3 particles accumulation}

\begin{figure}[!ht]
\centering
\includegraphics[width=7.5 cm]{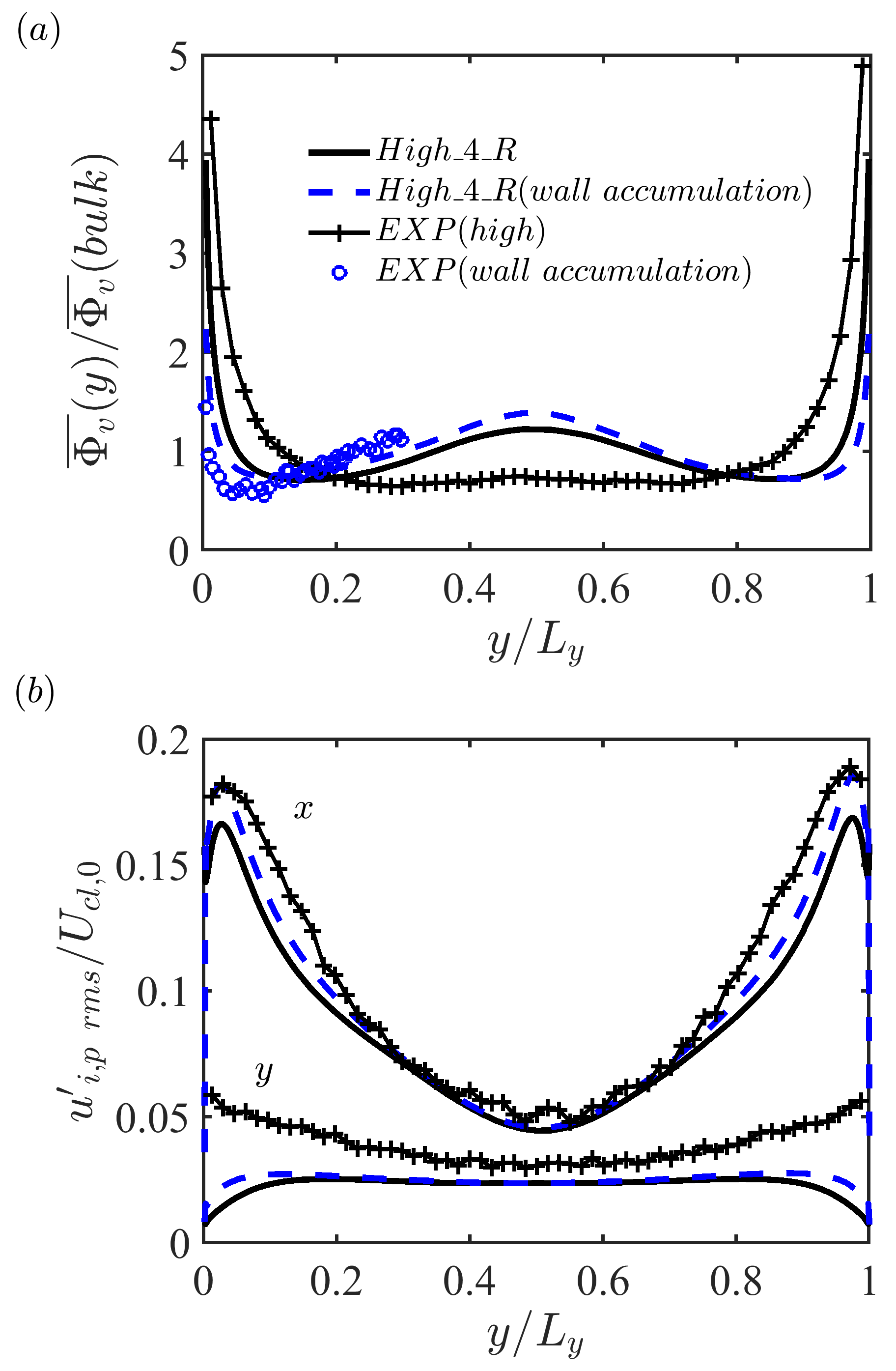} 
\caption{$(a)$ Particle concentration profiles and $(b)$ RMS particle fluctuation velocity, as a function of the wall-normal height. Case $high\_4\_R$ with both smooth wall and particle-induced roughness are compared with corresponding experimental observations of \cite{FongJFM2019}.}
\label{fig:particles accumulation}
\end{figure}

As reported by \cite{FongJFM2019}, when standard acrylic walls are used, the concentration profiles start with a strong near-wall peak but drift in time, with particles migrating away from the wall due to collsions with particles adhering to the walls and creating an effective roughness. Due to the roughness, \cite{FongJFM2019} observed more particles in the center region and fewer particles in the near-wall region, as shown in Fig. \ref{fig:particles accumulation}(a). Based on visual observation, the particle layer which forms with standard acrylic walls covers roughly $10-30\%$ of the wall surface. However, after replacing the standard acrylic walls with electrostatic dissipative acrylic walls, particles no longer adhere electrostatically to the wall.

Therefore as an additional test, we artificially place a particle layer at the wall, covering $30\%$ of the wall surface with randomly-located, fixed particles identical in size to the suspension (no two-way coupling feedback is included for these particles). The particle concentration and RMS particle fluctuation velocity are shown in Fig. \ref{fig:particles accumulation}(a) and (b), respectively, for high mass loading. Due to the presence of this particle-induced roughness, fewer particles are found close to the wall while more particles drift towards the channel center. Particle concentration profiles agree well between the numerical simulation and experimental observations using standard acrylic walls. The increased roughness consequently enhances the RMS particle fluctuation velocity close to the wall, which is consistent with previous investigations of the roughness, e.g., the simulations of \citep{vreman2015turbulence} and experiments of \cite{benson2005effects}. We note that magnitude of these roughness effects is weaker in present study than in previous investigations. This might be due to the fact that the present roughness includes the full particles mounted to the wall, while \cite{vreman2015turbulence} use smaller hemispheres. This results in collisions only between suspended particles with convex surfaces, and this has been seen qualitatively in additional tests (not shown here). 

\section{Comparison of particle statistics}
\label{4. Statistic analysis}

In this section, the comparison of statistics between the two- and four-way coupled model of \citet{richter2013momentum}, the four-way coupled model of \cite{capecelatro2013euler}, and the experimental observations of \cite{FongJFM2019} are shown in detail for both high and low mass loadings. This not only includes the more common statistical quantities including particle concentration profiles, first (i.e. mean velocity) and second order moments (i.e. turbulent kinetic energy and Reynolds shear stress) of the particle phase, but also particle statistics associated with fluid structures (i.e. skew of the particle and fluid fluctuation velocity, particle concentration dependence of low and high speed streaks), particle clustering behaviours analyzed by domain tessellation techniques (i.e. Vorono\"i diagram and box counting method), and two-particle statistics (i.e. radial distribution function and angular distribution function). All statistics are taken after a statistically steady state has been achieved (see Fig. \ref{fig:urms_evolution}).\\

\subsection{Particle concentration}
\label{4.1 Concentration}

\begin{figure}[!ht]
\centering
\includegraphics[width=7.5 cm]{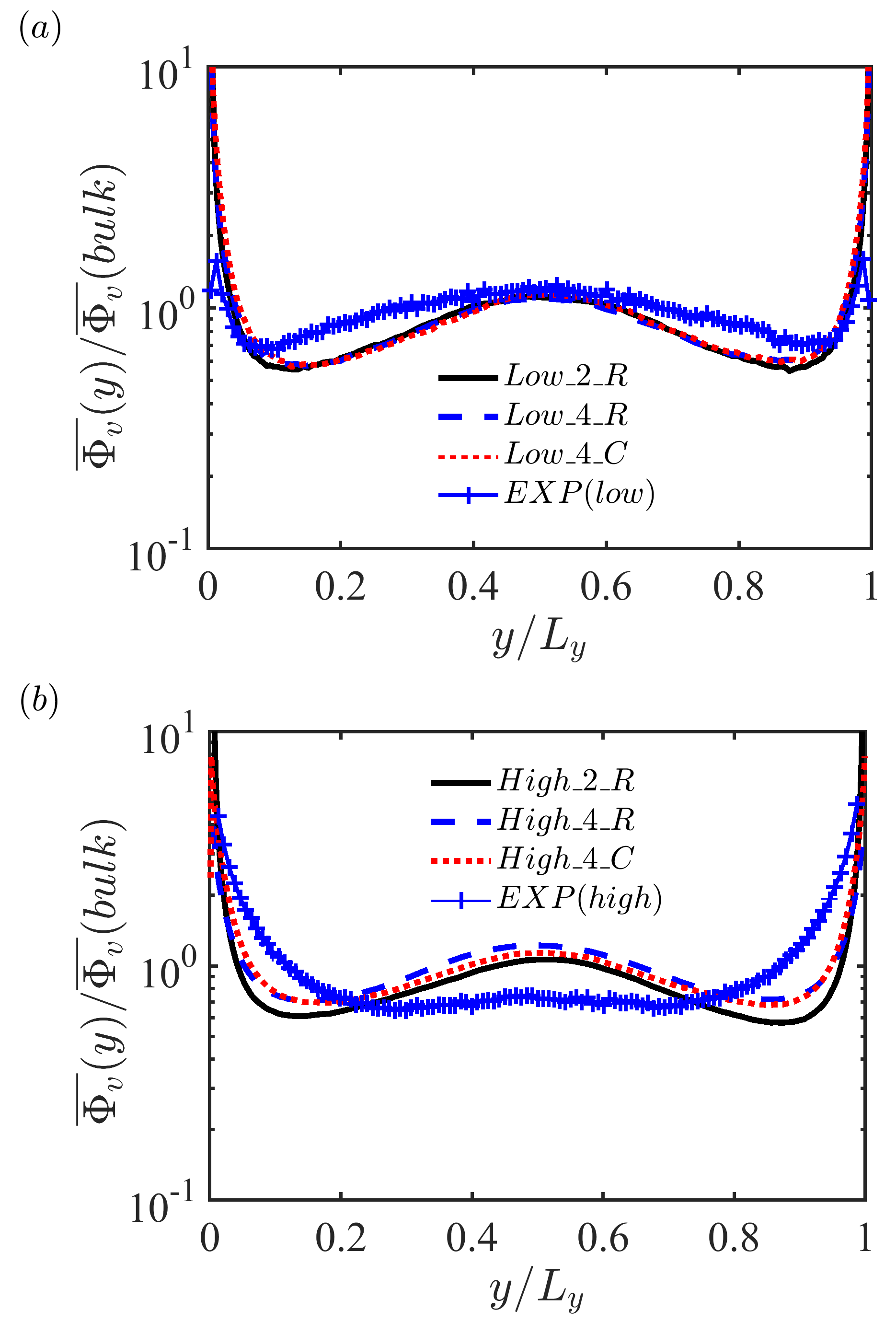} 
\caption{Particle concentration profile in wall-normal direction, comparison between simulations and the experiment: $(a)$ low concentration; $(b)$ high concentration.}
\label{fig:Concentration}
\end{figure}

The particle concentration profiles normalized by the bulk concentration are shown in Fig. \ref{fig:Concentration}. Based on the numerical simulations, low or high mass loading leads to similar particle distributions. With low mass loading as in Fig. \ref{fig:Concentration}(a), the profiles nearly overlap between the numerical simulations. In addition, the profile shape is similar between numerical simulations and the experimental measurements, while the influence of turbophoresis in the simulation is stronger at the wall than in the experiment. In both numerical simulations, the effect four-way coupling is minimal, suggesting that particle-particle collisions are not a dominant effect.\\

For high mass loading, the particle distribution exhibits a measurable difference when comparing between all numerical simulations with the experimental observations, as shown in Fig. \ref{fig:Concentration}(b). From the experiment, the concentration decreases monotonically in the wall-normal direction towards the channel centerline. There are more particles in the near-wall region in the case of high mass loading than low mass loading, which corresponds to an opposite trend in the channel center. Above, Fig. \ref{fig:VS_Jesse_2}(a) shows via numerical simulations that very high inertia particles ($St^+=2030$) with high mass loading ($\Phi_m=0.8$) tend to attenuate the turbulence \citep{vreman2015turbulence}, leading to a flatter concentration profile than with low mass loading ($\Phi_m=0.15$) \citep{capecelatro2015mass}. Even at these high mass loadings and despite a relatively flattened concentration profile flattened by high mass loading, there still exists a local maxmum in concentration near the channel center. Therefore the discrepancy between the experiments and simulations regarding the sensitivity of $\Phi_{V}(y)$ to the bulk mass loading remains unclear. But it seems to suggest that at this mass loading the particles are affecting the fluid flow in a way that modifies turbophoresis \citep{FongJFM2019}. We note that the logarithmic scale of the concentration amplifies these differences.\\

\begin{figure}[!ht]
\centering
\includegraphics[width=7.5 cm]{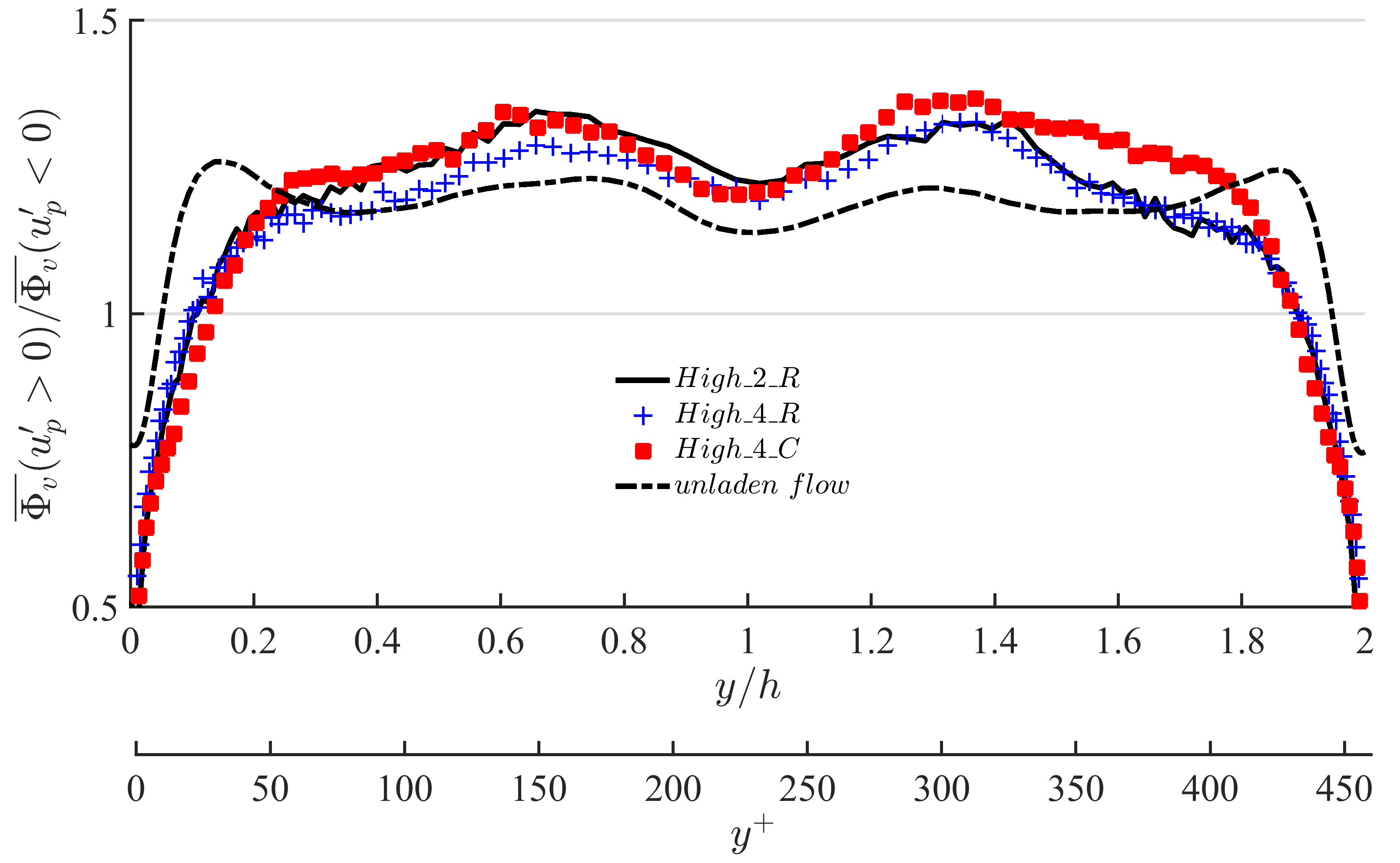} 
\caption{The ratio between particle concentration with $\overline{\Phi_v}(u'_p>0)$ and $\overline{\Phi_v}(u'_p<0)$. For unladen flow, the number ratio of Eulerian grid points with $u'_f>0$ or $u'_f<0$ are plotted.}
\label{fig:Concentration_posi_nega}
\end{figure}

Inertial particles preferentially accumulate in low-speed streaks --- a feature that has been previously observed by both experimental observations \citep{kaftori1995particle1} and numerical investigations \citep{Pan1996PoF, marchioli2002mechanisms, richter2013momentum, wang2019modulation, wang_richter_2019} in different configurations (e.g., channel flow, open channel flow or planar Couette flow) over a wide range of Reynolds numbers ($Re_\tau=40-1000$). In a horizontal open channel flow, \cite{sumer1981particle} experimentally found that heavy particles near the bottom are swept into low-speed wall streaks, from whence they are ejected again into the flow. This phenomenon is also observed in current configuration. From the simulation data, we count the particles with $u'_p<0$ or $u'_p>0$ to represent particle numbers in low or high speed regions, where $u'_p$ is the particle fluctuation velocity at a particular wall-normal distance. The ratio of $\overline{\Phi_v}(u'_p>0)$ with $\overline{\Phi_v}(u'_p<0)$, cast in terms of the effective volume concentration corresponding to these particle counts, is shown in Fig. \ref{fig:Concentration_posi_nega}. Simultaneously, Eulerian grid points with $u'_f>0$ or $u'_f<0$ are plotted for the unladen flow, where $u'_f$ is the fluid fluctuation velocity. Across the channel, computed results from the numerical simulations agree well with each other. Close to the wall, more particles are in low speed regions than in high speed regions which is opposite compared to the center region. In the near-wall region, this can be explained by the mechanism proposed by \cite{sumer1981particle} noted above, that heavy particles near the bottom are swept into low-speed wall streaks, from where they are ejected again into the flow. In the core region of the channel, the preferential sweeping mechanism for a heavy particle interacting with local flow vortical structures under its inertia and the streamwise gravity proposed by \cite{wang1993settling} is a possible explanation for this.  

\subsection{Flow and particle velocity statistics}
\label{4.2 Velocity}

\subsubsection{Mean velocity}
\label{4.2.1 Velocity fluctuation}

\begin{figure*}[!ht]
\centering
\includegraphics[width=16 cm]{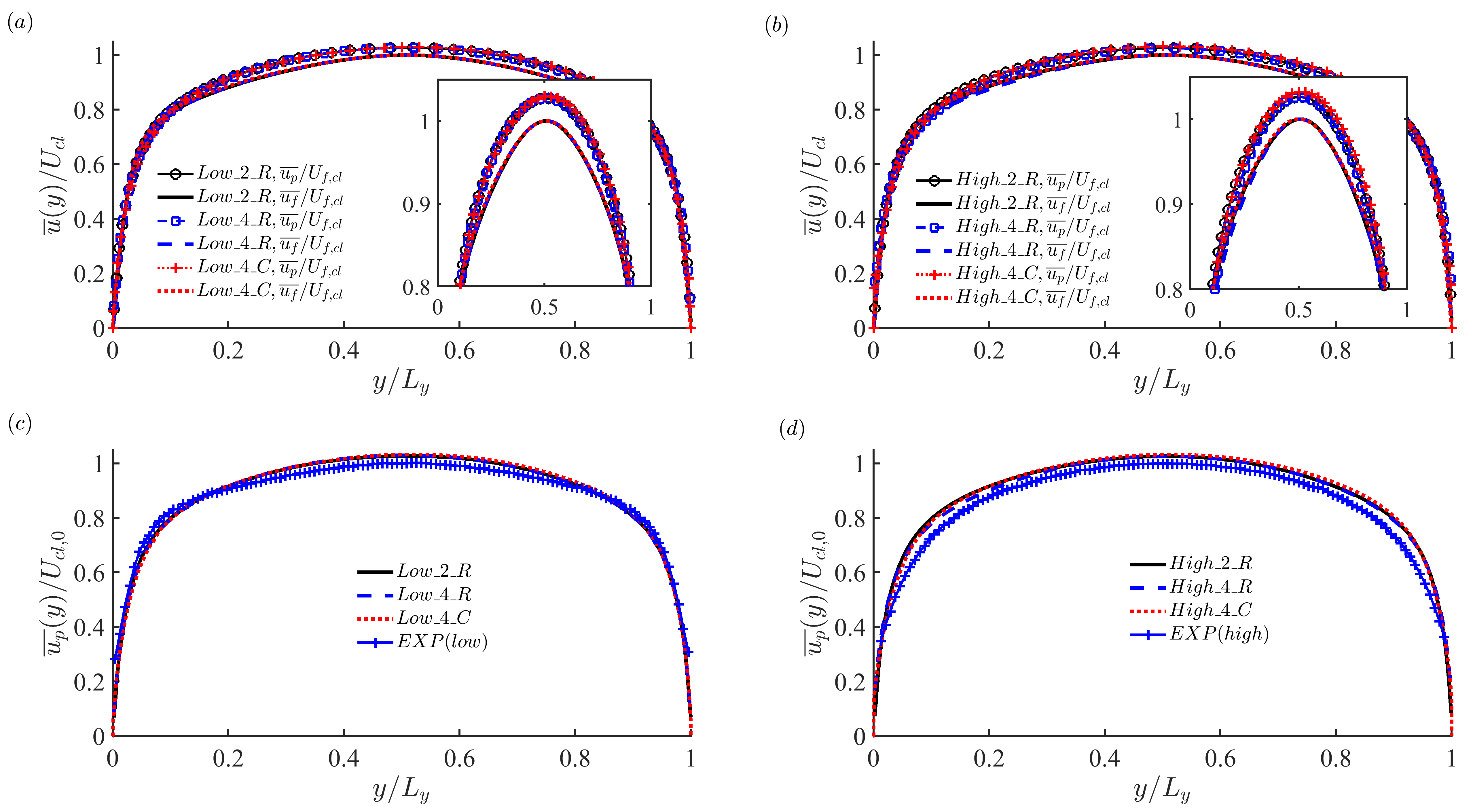} 
\caption{Mean streamwise velocity in wall-normal direction: $(a,c)$ low mass loading; $(b,d)$ high mass loading. $(a,b)$ Particle phase and carrier phase scaled by the fluid centerline velocity in particle-laden flow; $(c,d)$ Particle phase normalized by the fluid centerline velocity in unladen flow.}
\label{fig:U_ave}
\end{figure*}

Mean velocity profiles of the particle phase and carrier phase are shown in Fig. \ref{fig:U_ave}, where panels (a,c) are for the low mass loading case and (b,d) are for the high mass loading case. In Figs. \ref{fig:U_ave}(a,b), the mean velocity of the particle phase ($\overline{u_p}$) and fluid phase ($\overline{u}$) are shown from numerical simulations, and are normalized by the fluid centerline velocity of the particle-laden flow. Both $\overline{u_p}$ and $\overline{u}$ overlap between the two-way coupling and four-way coupling simulations, indicating that in this dilute limit both two- and four-way coupling have weak impacts on the mean flow. The fluid velocity ($\overline{u}$) lags slightly behind the particle velocity ($\overline{u_p}$) in the majority region of the channel ($0.15<y/L_y<0.85$), which is also observed by \cite{capecelatro2015mass} and \cite{benson2005effects} in vertical channel flow but for a higher inertia particles ($St^+=2030$). As previously shown in Fig. \ref{fig:Concentration_posi_nega}, more particles in high-speed regions away from the wall result in a higher particle average velocity compared to the fluid mean velocity\\

In Figs. \ref{fig:U_ave}(c,d), we compare mean velocity of the particle phase ($\overline{u_p}$) between numerical simulations and experimental observations, normalized by the fluid centerline velocity of unladen flow. Numerical simulations give an almost identical results between low and high mass loading. However, compared with the numerical simulations, the experimentally observed particle average velocity profile is slightly flatter in the case of low mass loading but more parabolic with high mass loading. This higher sensitivity to mass fraction in the experiments as compared to the simulations is similar to that shown in the concentration profiles, but the differences are small ($\Delta \overline{u_p}/\overline{u_p}(EXP)<5\%$). 

\subsubsection{Particle velocity fluctuations}
\label{4.2.2 Velocity fluctuation}

\begin{figure*}[!ht]
\centering
\includegraphics[width=16 cm]{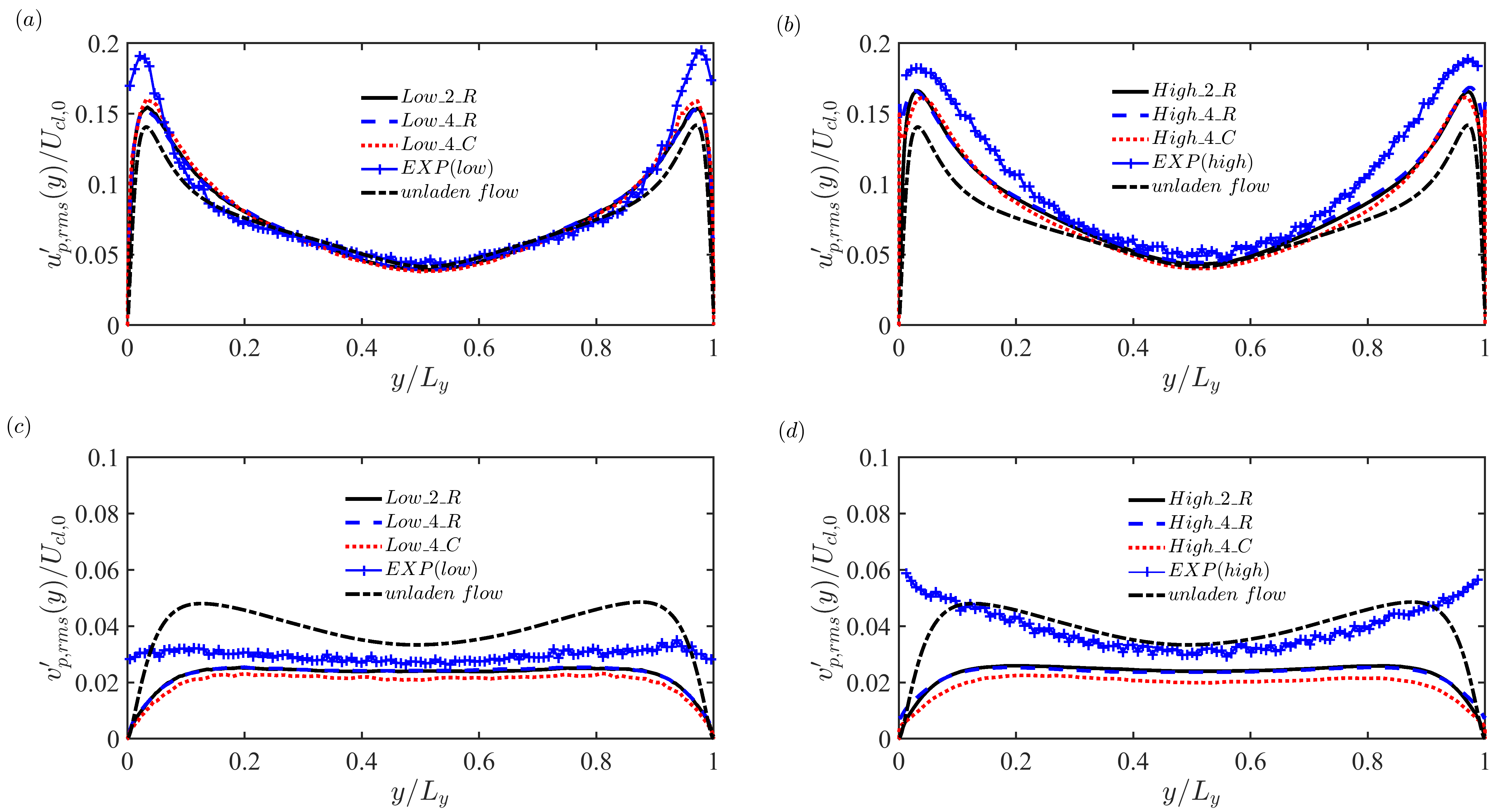}
\caption{RMS particle fluctuation velocity in wall-normal direction, normalized by the fluid centerline velocity in unladen flow: $(a,c)$ low mass loading; $(b,d)$ high mass loading. $(a,b)$ RMS particle streamwise fluctuation velocity, $u'_{p,rms}$; $(c,d)$ RMS particle wall-normal fluctuation velocity, $v'_{p,rms}$. For comparison, $u'_{rms}$ and $v'_{rms}$ in the simulated case of $Unladen\_R$ are plotted.}
\label{fig:uvw_rms}
\end{figure*}

Figs. \ref{fig:uvw_rms}(a,b) and (c,d) show the RMS particle fluctuation velocity in the streamwise and wall-normal directions, respectively. In the streamwise direction, the particle RMS fluctuation velocity is higher than the fluid's close to the wall, especially in the experiment. This is similar to the comparison between the numerical simulations of \cite{capecelatro2015mass} with experimental measurements from \cite{benson2005effects} for high inertia particles ($St^+=2030$). Away from the wall, numerical simulations correspond to the measurements closely. For the wall-normal component of the RMS fluctuation velocity, $v'_{p,rms}$ is smaller than the fluid's across the entire channel, again similar to the experiments of \cite{kulick1994particle} for a wide range of Stokes numbers ($St^+=400-2030$). Close to the wall, the measured profile in \cite{FongJFM2019} remains fairly flat across the channel and largely exceeds the unladen fluid levels and it does not appear to vanish. At high mass loading, this difference becomes more enhanced. Again, the simulations exhibit very similar behavior at both mass loadings, while the experiments see an increase in wall-normal particle velocity fluctuations. \\

\begin{figure}[!ht]
\centering
\includegraphics[width=7.0 cm]{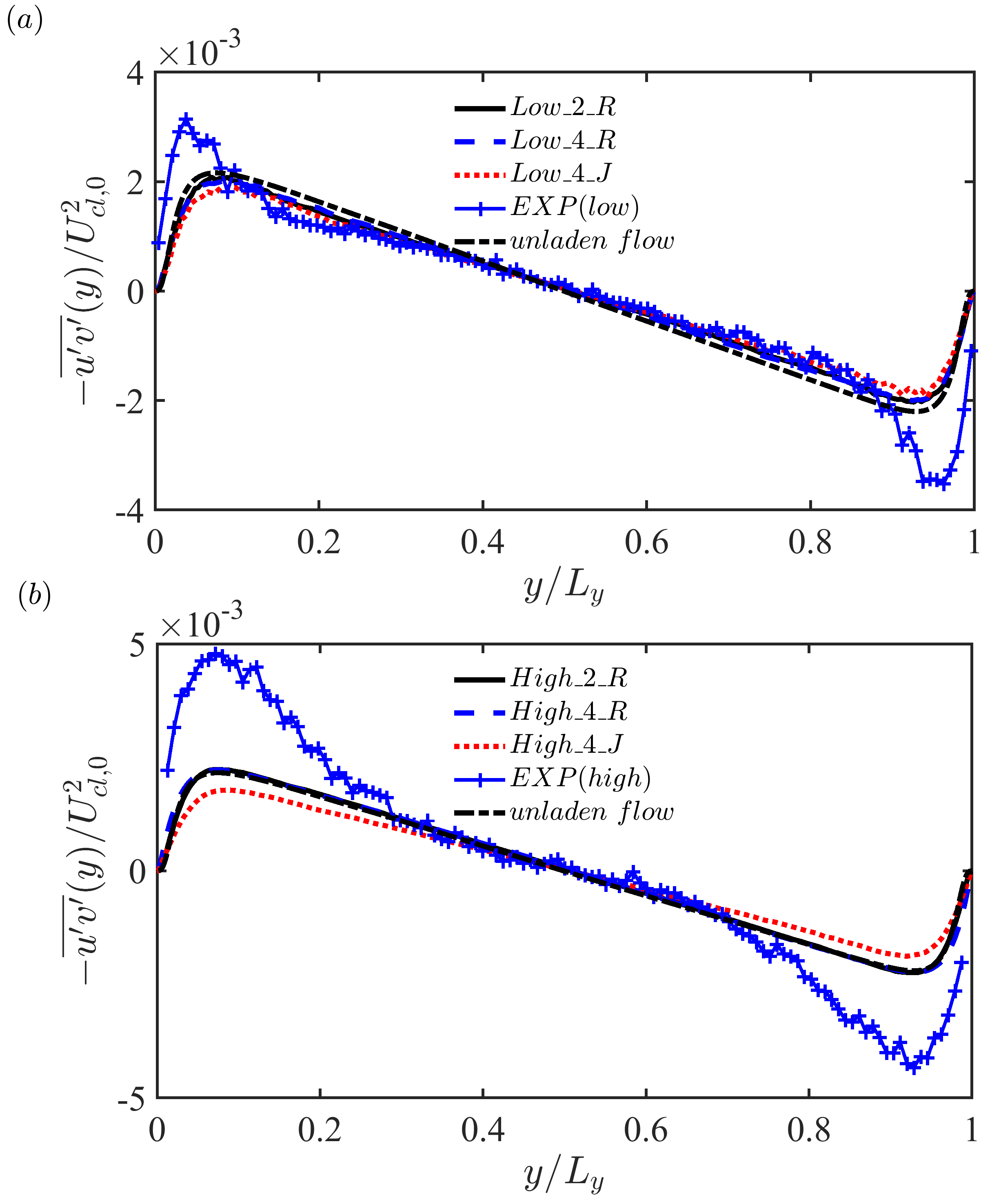}
\caption{Particle Reynolds shear stress as a function of wall-normal direction normalized by the fluid centerline velocity in unladen flow: $(a)$ low mass loading; $(b)$ high mass loading. For comparison, $u'v'$ in the case of Unladen\_R is plotted.}
\label{fig:uv_particle}
\end{figure}

The enhanced particle wall-normal fluctuation measured in the experiments consequently contributes to a higher particle Reynolds shear stress as shown in Fig. \ref{fig:uv_particle}. The discrepancy between numerical simulations with the experiment appears largely confined to the inner layer ($y/L_y<0.2$ corresponding to $y^+<90$) with high mass loading, while they nearly overlap with each other in the channel center (i.e. outer layer $y^+>100$).\\

\begin{figure}[!ht]
\centering
\includegraphics[width=7.0 cm]{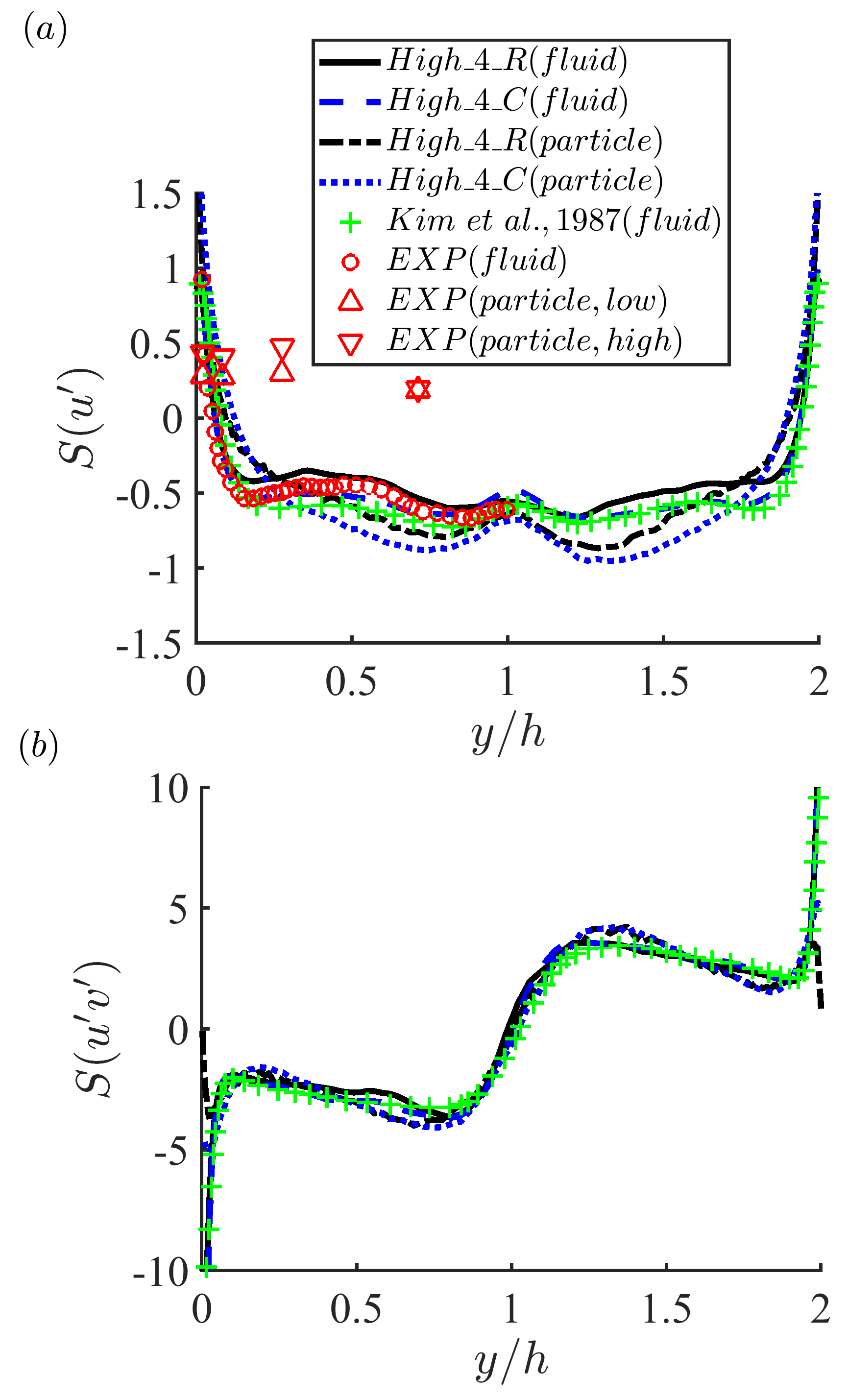}
\caption{Skewness factor of both particle and fluid phases in the case of high mass loading: $(a)$ $S(u')$ and $(b)$ $S(u'v')$. As a comparison, $S(u')$ and $S(u'v')$ of the single-phase flow from \cite{kim1987turbulence} at $Re_\tau=180$ are plotted.}
\label{fig:skewness}
\end{figure}

The computed skewness factor $S(u')$ of the fluctuating velocity distribution ($u'_p$ for the particle and $u'$ for the carrier phase in Eulerian grid) and $S(u'v')$ ($u'_p v'_p$ for the particle and $u' v'$ for the carrier phase) is shown in Fig. \ref{fig:skewness}. Here, negative skewness indicates a heavy tail occurring on the left side of the distribution, and positive skewness indicates that a heavy tail occurs on the right. The skewness factor in Fig. \ref{fig:skewness}(a) for $u_p'$ and $u'$ is also compared with previous DNS data for single-phase flow at $Re_\tau=180$ computed by \cite{kim1987turbulence}. For the carrier phase ($u'$), there is general agreement between all simulation results (current simulations and \cite{kim1987turbulence}) and measured data, showing a positive skewness factor close to the wall but negative away from the wall, with a crossover point at $y^+=20 \sim 30$. In the current numerical simulations, the behaviour of the skew of $u'_p$ is similar, and the crossover point moves to a higher $y^+=50 \sim 65$ compared that for $u'$. This is consistent with the discussion of Fig. \ref{fig:Concentration_posi_nega}, that particles reside more in the low-speed streaks close to the wall but in high-speed streaks away from the wall. Compared to the measured values of $S(u'_p)$, however, a different behavior is seen. The symbols in Fig. \ref{fig:skewness}(a) indicate that the probability distribution of $u'_p$ has a similar shape across the channel, and therefore the skew of $u'_p$ has an opposite sign with $u'$ away from the wall. This is in contrast to the simulations, which show a similar qualitative behavior between $S(u')$ and $S(u'_p)$.\\

The skewness of Reynolds shear stress $u'v'$ is shown in Fig. \ref{fig:skewness}(b), which is antisymmetric about the center plane. From the wall to channel center, a negative skewness factor indicates that the tail is always on the left side of the probability distribution of $u'v'$. In single-phase flow, the main contribution of the Reynolds shear stress is from ejections and sweeps in wall-bounded turbulence \citep{kim1987turbulence}. Consequently, the majority of particle Reynolds shear stress is due to particles in the same ejections and sweeps. In the experiment of \cite{FongJFM2019}, it is not possible to check whether $u'_p$ and $u'_f$ or $v'_p$ and $v'_f$ have the same sign at the particle positions due to the lack of simultaneous carrier phase measurements. However they observed that the majority of the particle Reynolds shear stress is due to the contribution from the second and fourth quadrants of the $(u'_p,v'_p)$ plane.

\subsection{Domain tessellation}
\label{4.3 Tessellation}

Particle preferential accumulation is a key feature of inertial particle behavior and has been shown to have a significant impact on turbulence modification \citep{eaton1994preferential}. When combined with gravitational settling, inertia has been seen to influence the effective settling rate, where particles can fall at speeds not equal to their terminal velocity \citep{petersen2019experimental,wang1993settling}; this effect is tightly linked to preferential accumulation. Therefore, in this section we show particle clustering behaviours in the current study, and compare them to the results of \cite{FongJFM2019}. There are several techniques to identify clusters of particles, and almost all the methods try to quantify the deviation from a uniform distribution of particles \citep{monchaux2012analyzing}. Following \cite{FongJFM2019}, we focus on Vorono\"i tessellation and the box counting method in this section.\\

\subsubsection{Vorono\"i tessellation}
\label{4.3.1 Voronoi Tessellation}
\begin{figure*}[!ht]
\centering
\includegraphics[width=15 cm]{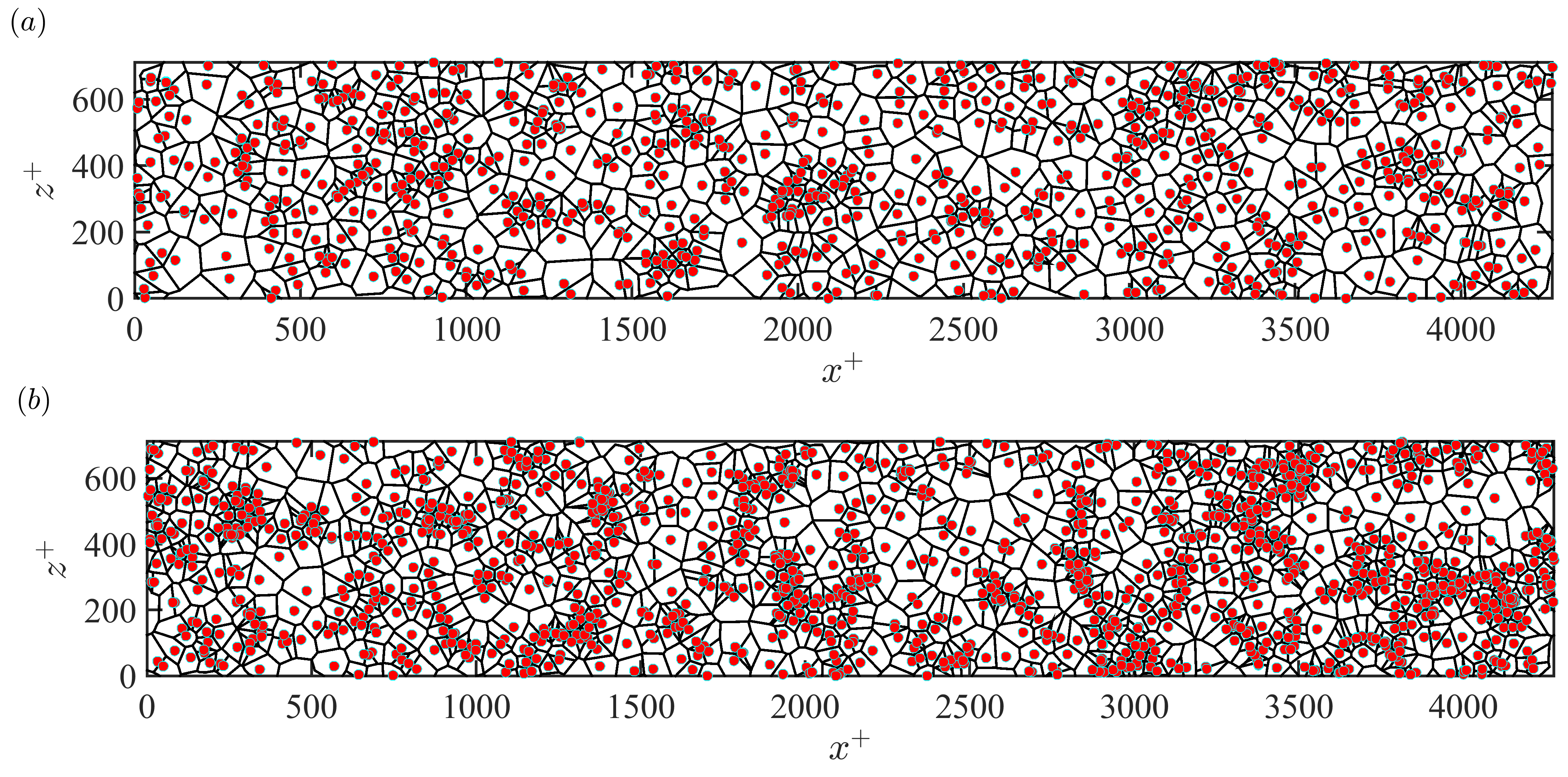}
\caption{Particle locations and the associated Vorono\"i diagram in a wall-normal slab with thickness of $2 d_{p}$ at two wall-normal locations: $(a)$ near-wall region $y/L_y=0.1$ ($y^+=46$); $(b)$ channel center $y/L_y=0.5$  ($y^+=227$).}
\label{fig:Voronoi_distribution}
\end{figure*}

Fig. \ref{fig:Voronoi_distribution} shows instantaneous particle locations and the associated Vorono\"i diagram in a slab with thickness of $2d_p$ close to the wall ($y/L_y=0.1$) and in the center region ($y/L_y=0.5$) of case $high\_4\_R$. Based on the particle concentration profiles shown in Fig. \ref{fig:Concentration}, there are less particles in the former ($y/L_y=0.1$) than in the latter ($y/L_y=0.5$), which can be observed in the particle distributions in Figs. \ref{fig:Voronoi_distribution}(a) and (b). \\

\begin{figure}[!ht]
\centering
\includegraphics[width=7.5 cm]{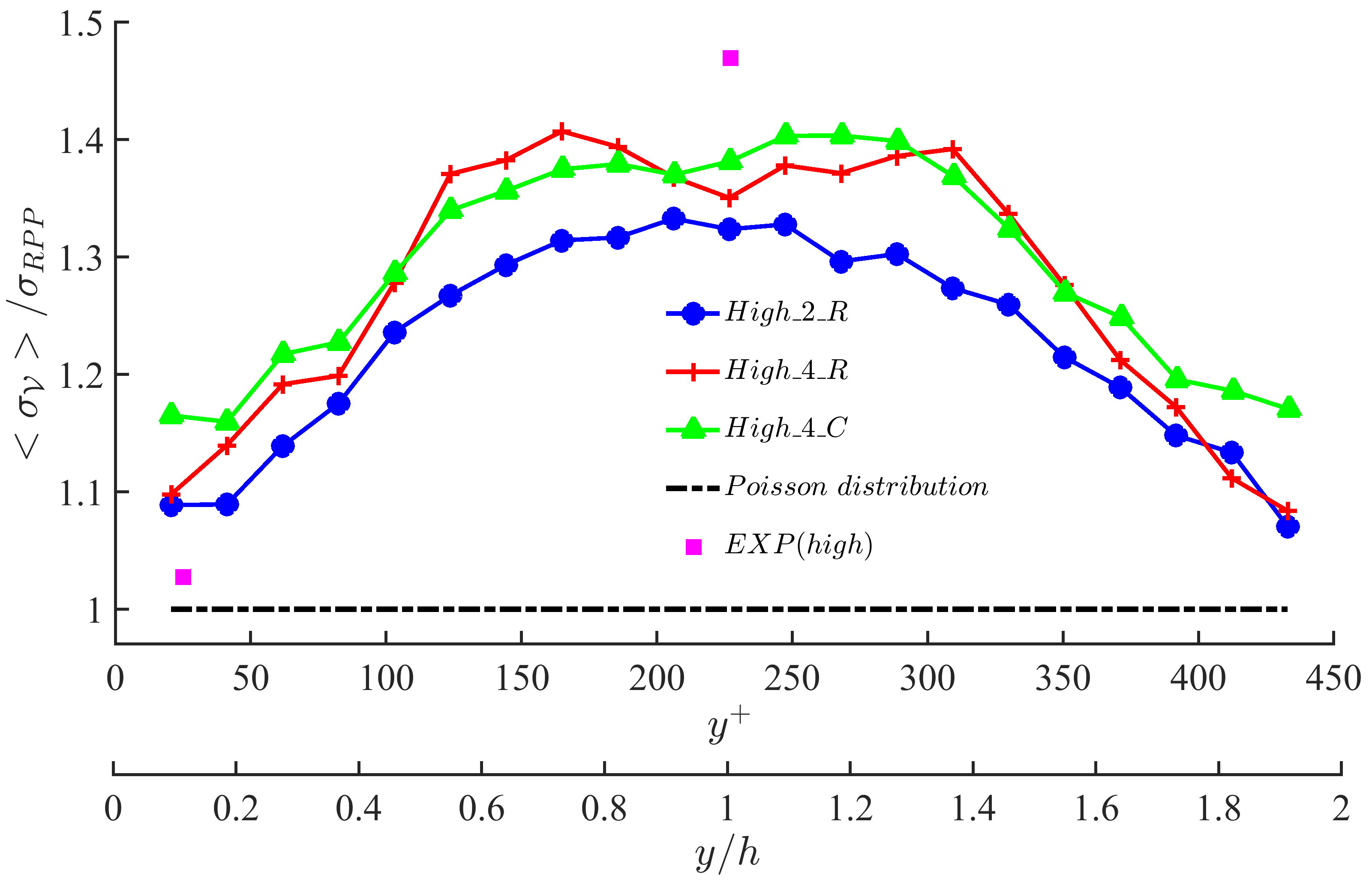}
\caption{Standard deviation of the normalized Vorono\"i area $\sigma_{\mathcal{V}}$ in a wall-normal slab with thickness of $2 d_{p}$, normalized by that of a random Poisson process, $\sigma_{RPP}$ as a function of height in wall-normal direction of the case with high mass loading. Experimental observation from from \cite{FongJFM2019} is shown as a black square close to the wall ($y^+=25$) and at the channel centerline.}
\label{fig:Voronoi_variance}
\end{figure}

In order to quantify the particle clustering behaviour, we employ a Vorono\"i diagram analysis, which compares the distribution of the tessellation areas in the particle-laden cases with the expected random Poisson process (RPP) if the particles were uniformly distributed (see for example \cite{monchaux2012analyzing}). Fig. \ref{fig:Voronoi_variance} shows the standard deviation ($\sigma_\mathcal{V}$) of the distribution of the normalized Vorono\"i area $\mathcal{V}=A/\overline{A}$, where the inverse of the average Vorono\"i area $\overline{A}$ indicates the mean particle concentration. $\sigma_\mathcal{V}$ is scaled by the standard deviation of a random Poisson process (RPP; $\sigma_{RPP}=0.52$). A ratio $\sigma_\mathcal{V}/\sigma_{RPP}$ exceeding unity indicates that particles are accumulating in clusters as compared to truly randomly distributed particles. \\

Fig. \ref{fig:Voronoi_variance} shows the ratio $\sigma_\mathcal{V}/\sigma_{RPP}$ for multiple wall-normal distances. Across the entire channel, computed results from the two-way coupling configuration is slightly lower than those from both of four-way coupling methods, which indicates that particle preferential accumulation is higher when collision forces are included. This is in contrast with previous studies without gravity by \cite{li2001numerical} and \cite{nasr2009dns}, who found that particle/particle collisions weakened the preferential distribution of particles. Comparing the two four-way coupling cases, the standard deviation is higher for case $high\_4\_C$ than for $high\_4\_R$ near the wall; aside from this, they have a good agreement with each other away from the wall ($y/h>0.11$ or $y^+>40$). In addition, the ratio $\sigma_\mathcal{V}/\sigma_{RPP}$ increases monotonically with increasing wall-normal distance (towards the center), which indicates that the particle clustering effect is stronger in the center region but weaker in the near-wall region. The measured $\sigma_\mathcal{V}/\sigma_{RPP}$ of \cite{FongJFM2019} is from particles in slabs with thickness of $17$ viscous units. We can see that the measured $\sigma_\mathcal{V}/\sigma_{RPP}$ is lower close to the wall whereas higher in the center compared to the numerical simulations.

\subsubsection{Box counting method}
\label{4.3.2 Box counting}

\begin{figure*}[!ht]
\centering
\includegraphics[width=16 cm]{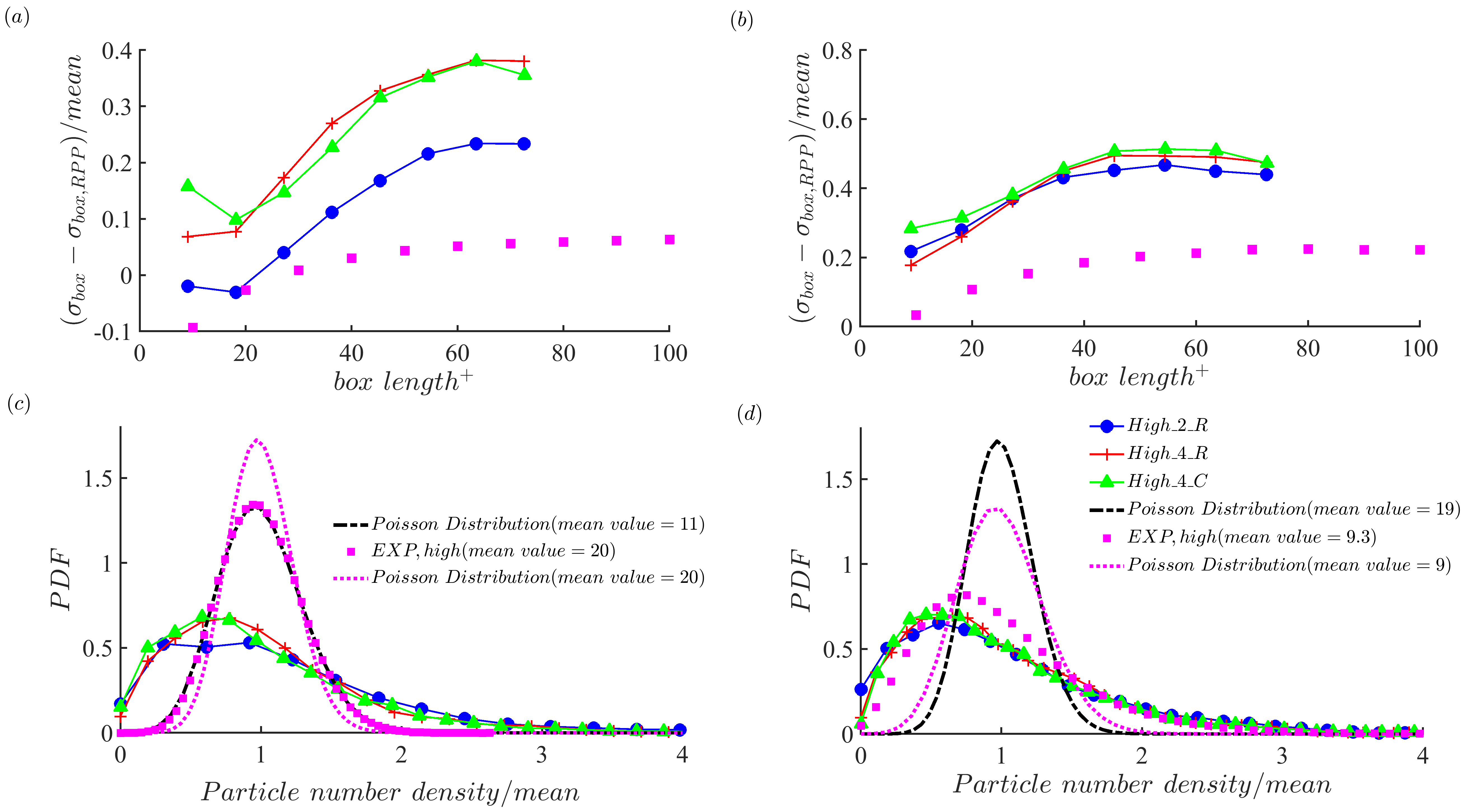}
\caption{(a,b) Deviation of particle number density distributions from uniformly distributed particles in a slab, as a function of the length of a square box. (c,d) distributions of particle number density within the slab. Slabs ($\Delta x^+=60$ and $\Delta z^+=60$) with the thickness of $\Delta y/L_y=0.036$ ($\Delta y^+=17$) are taken at two different wall-normal heights: (a,c) $y/L_y=0.12$ ($y^+=55$); (b,d) $y/L_y=0.5$ ($y^+=227$). Mean value represents for particle number per box.}
\label{fig:box_fix}
\end{figure*}

The Vorono\"i diagram analysis compares the distribution of the tessellation areas of the particles with the randomly distributed particles. Besides the Vorono\"i diagram analysis, here we use the box counting method to compare the particle number in a square box with that for randomly distributed particles; this process explores the length scale on which preferential concentration occurs. Thus the deviation of particle number density should be the same for randomly distributed particles when very small or large boxes are used to count for particle number density; see \cite{fessler1994preferential} and \cite{monchaux2012analyzing}. \\

The box size should be determined in order to capture the maximum deviation from the uniform particle distribution, which can be expressed as $(\sigma_{box}-\sigma_{box,RPP})$ scaled by the mean particle number density. The difference of $(\sigma_{box}-\sigma_{box,RPP})$ is shown in Figs. \ref{fig:box_fix}(a) and (b), which is computed in the near-wall region ($y/L_y=0.12$) and in the channel center ($y/L_y=0.5$), respectively. Generally, the deviation is larger in the center than that close to the wall, similar to the deviation of the Vorono\"i diagram analysis in Fig. \ref{fig:Voronoi_variance}. Specifically, in the near-wall region as in Fig. \ref{fig:box_fix}(a), the computed deviation with the use of the two-way coupling method is smaller than the four-way coupling configurations;  again this has the same trend as from the Vorono\"i diagram analysis shown in Fig. \ref{fig:Voronoi_variance}. In both the near-wall and center regions (Fig. \ref{fig:box_fix}(a) and (b), respectively), numerical simulations have similar deviations, however, the magnitudes are higher than the measurements of \cite{FongJFM2019}. Additionally, the maximum deviation occurs on box length scales of roughly $60$ wall units, which is similar to the experimental result from \cite{fessler1994preferential}.\\

Consequently, particle number density distributions calculated by a box size of $60$ wall units ($0.133L_y$) are shown in Figs. \ref{fig:box_fix}(c) and (d) for the near-wall region ($y/L_y=0.12$) and in the channel center ($y/L_y=0.5$), respectively. The distributions computed by numerical simulations agree well with each other. In the near wall region as shown in Fig. \ref{fig:box_fix}(c), the distributions of moderate inertia particles ($St^+=58.6$) are different from the randomly distributed particles, even though the deviation of the Vorono\"i diagram analysis in Fig. \ref{fig:Voronoi_variance} is as low as $1.1 \sim 1.2$. The measurements of \cite{FongJFM2019} show more particles in the box (mean value of $20$) than the numerical simulations (mean value of $11$), which corresponds to the higher concentration profile in the experiment than in the numerical simulations in the near-wall region (as shown in Fig. \ref{fig:Concentration}(b)). The experiment results actually exhibit a near-Poisson distribution which indicates a more uniformly distributed particle layer close to the wall than that from the numerical simulations. In the center region shown in Fig. \ref{fig:box_fix}(d), the measured particle number density distribution by \cite{FongJFM2019} is similar to the current numerical simulations, while there are less particle numbers in the box (mean value of $9.3$) than in the numerical simulations (mean value of $19$), which corresponds to the lower concentration profile in the experiment than in the numerical simulations in the near-wall region (as shown in Fig. \ref{fig:Concentration}(b)). Both the numerical and experimental results show a different distribution from randomly distributed particles, indicating a measurable particle clustering effect in the channel center.

\subsection{Two-point statistics: radial and angular distribution function}
\label{4.4 RDF and ADF}

\begin{figure*}[!ht]
\centering
\includegraphics[width=15 cm]{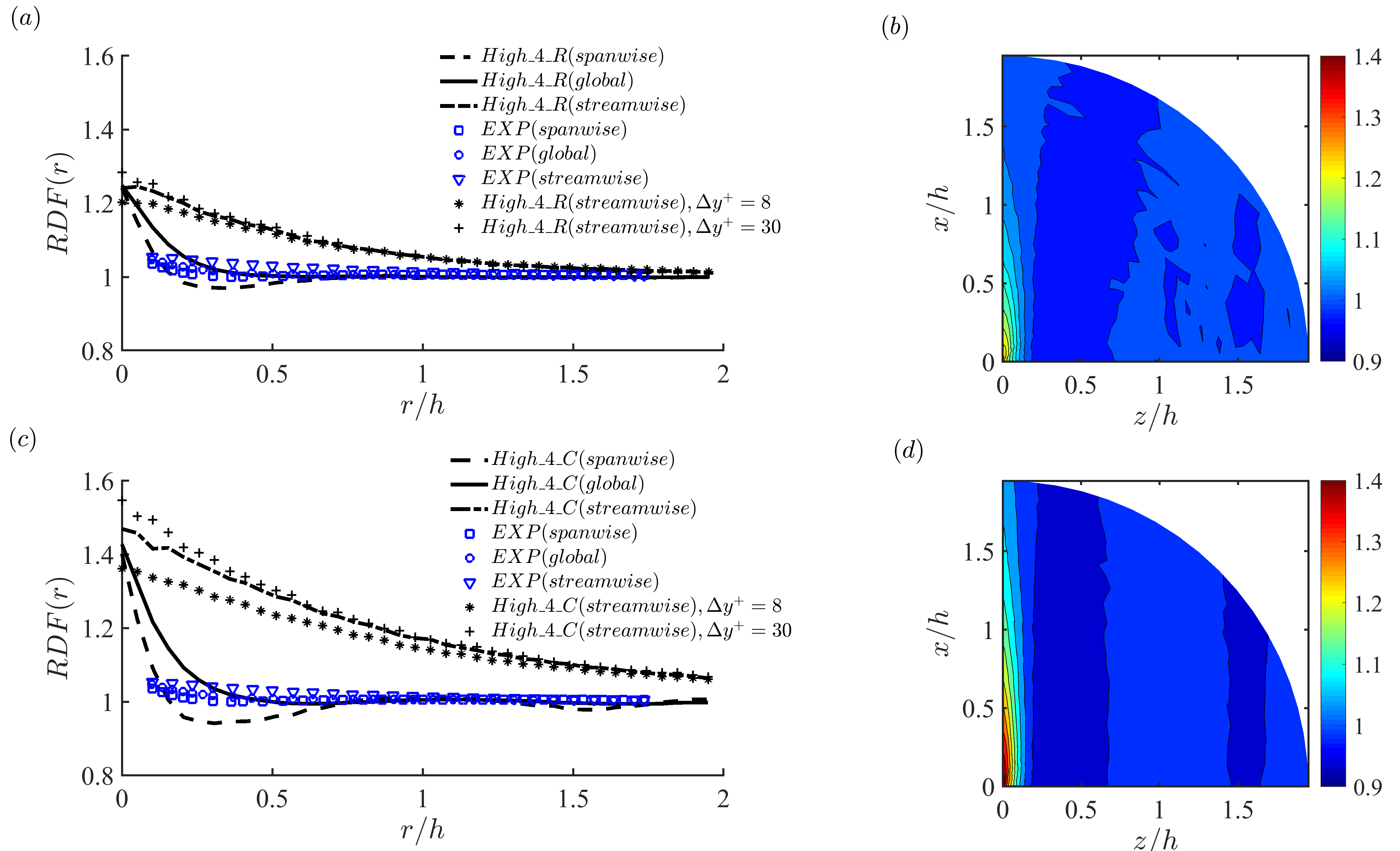}
\caption{RDF and ADF of particles in a slab with thickness of $\Delta y^+=17$ close to the wall ($y=0.11h-8.5\delta_ \nu$ to $y=0.11h+8.5\delta_ \nu$) in the case of high mass loading: (a,c) Global RDF, streamwise RDF (equals to $ADF(r,\theta=\pi/2)$) and spanwise RDF (equals to $ADF(r,\theta=0)$). Streamwise RDF in a slab with thickness of $\Delta y^+=8$ and $30$ is shown as a comparison, where spanwise and global RDF is similar between $\Delta y^+=17$ with $\Delta y^+=8$ and $30$; (b,d) Contours of ADF in the $x-z$ plane. (a,b) Case $high\_4\_R$; (c,d) Case $high\_4\_C$.}
\label{fig:RDF_wall}
\end{figure*}

\begin{figure*}[!ht]
\centering
\includegraphics[width=15 cm]{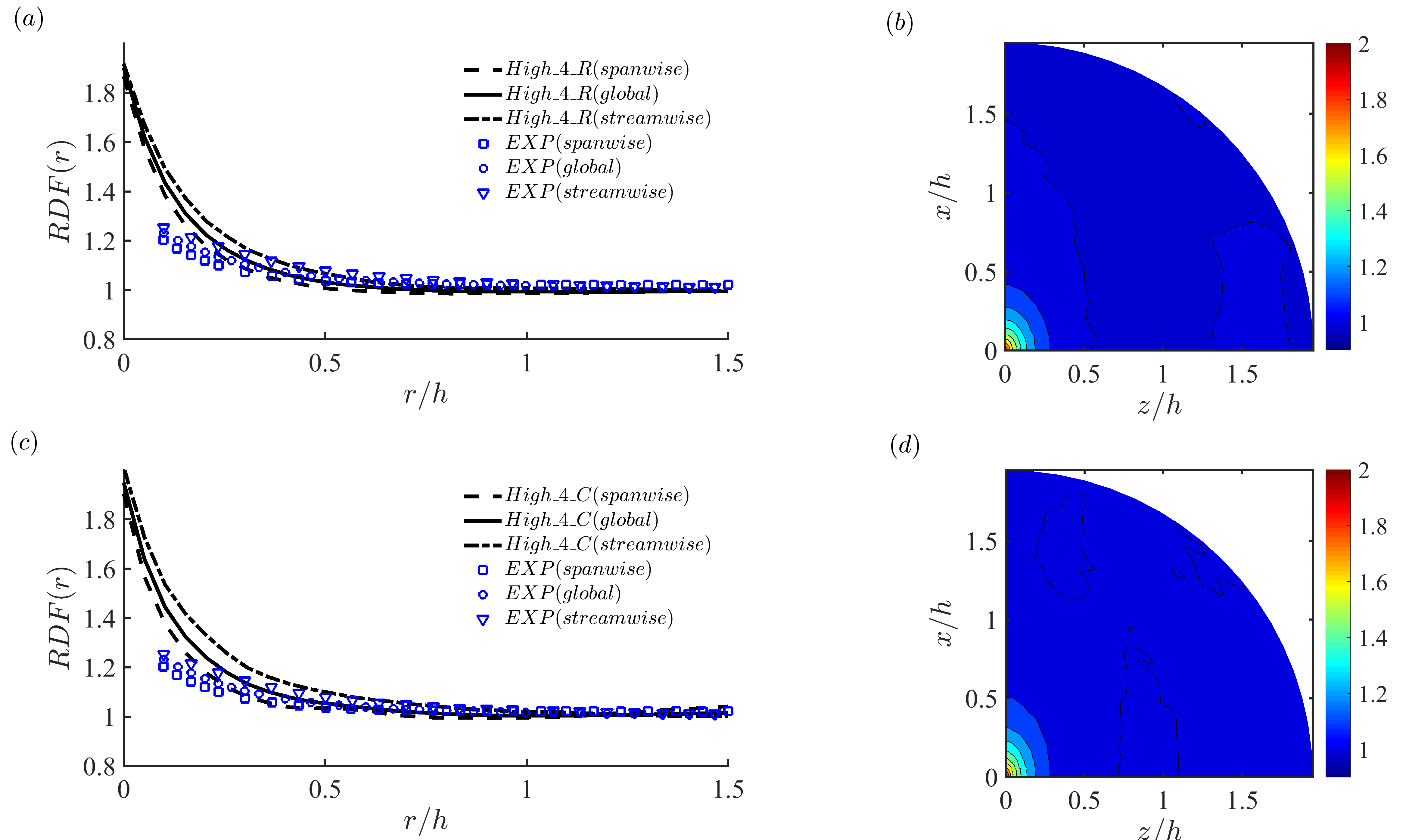}
\caption{RDF and ADF of particles in a slab with thickness of $\Delta y ^+=17$ along the center plane ($y=h-8.5\delta_ \nu$ to $y=h+8.5\delta_ \nu$) in the case of high mass loading: (a,c) Global RDF, streamwise RDF (equals to $ADF(r,\theta=\pi/2)$) and spanwise RDF (equals to $ADF(r,\theta=0)$); (b,d) Contours of ADF in the $x-z$ plane. (a,b) Case $high\_4\_R$; (c,d) Case $high\_4\_C$.}
\label{fig:RDF_center}
\end{figure*}

The Vorono\"i diagram analysis and box counting method provide a global metric of clustering without consideration of the cluster anisotropy that frequently appears in the near-wall region. In this regard, another widely used tool to quantify particle clustering is the radial distribution function (RDF). The radial distribution function describes how density varies as a function of distance from a reference particle, which has been successfully applied in homogeneous turbulence to determine particle clusters and collision mechanisms, see for example \cite{sundaram1997collision}, \cite{Barkley2007}, \cite{gualtieri2009anisotropic}, and \cite{bragg2015mechanisms}. However, the RDF still provides only an omni-directional average of particle information \citep{baker2017coherent}. Thus, an expansion of RDF from 1D to 2D polar coordinates, providing an angular distribution function (ADF), illustrates the anisotropy of particle clustering behaviour in both distance and direction \citep{gualtieri2009anisotropic,FongJFM2019}.\\

The two-dimensional radial and angular distribution functions are defined as in Eqs \ref{eq:RDF} and \ref{eq:ADF}, where in the simulations particles are taken from a slab with thickness of $0.036L_y$ ($17$ viscous units): 

\begin{equation}\label{eq:RDF}
RDF(r)= \frac{\sum_{i=1}^{n_p} \delta N_i(r)/ (\delta r \cdot n_p)}{N / (L_x \cdot L_y)},
\end{equation} 

\begin{equation}\label{eq:ADF}
ADF(r,\theta)= \frac{\sum_{i=1}^{n_p} \delta N_i(r,\theta)/ (\delta r \cdot \delta \theta \cdot  n_p)}{N / (L_x \cdot L_y)},0 \leq \theta \leq \pi/2,
\end{equation} 

\noindent where $\delta N_i(r)$ is the particle number between $r - \delta r /2$ and $r + \delta r /2$ from the center of particle $i$, and $\delta N_i(r,\theta)$ is the particle number in a sector between $r - \delta r /2$ and $r + \delta r /2$ in the radial direction and $\theta - \delta \theta /2$ and $\theta + \delta \theta /2$ in the angular direction from the center of particle $i$; $\theta=0$ and $\theta=\pi/2$ correspond to the spanwise and streamwise directions, respectively. In the present study, we set $\delta r=0.05h$ ($\delta r^+=11.4$) and $\delta \theta=0.025\pi$ to compute $RDF(r)$ and $ADF(r,\theta)$. The mean value is from the average of $n_p$ particles from multiple snapshots in time. Finally, the distribution functions are normalized by the surface average particle number in $x-z$ plane ($n_p / L_x L_y$ representing a randomly distributed particle number density), where $n_p$ particles are from a two-dimensional $x-z$ slab taken in the wall-normal direction. Periodic boundary conditions are used for particles near the boundaries in the streamwise and spanwise directions. \\

Close to the wall, particle preferential accumulation in streaky structures has been observed both numerically (e.g. \cite{rouson2001preferential}) and experimentally (e.g. \cite{kaftori1995particle1}). The clustering of particles preferably appears for moderate Stokes number particles, i.e. $St^+=O(10)$ \citep{wang2019modulation}. The RDF and ADF of particles in a slab close to the wall ($y=0.11h-8.5 \delta_ \nu$ to $y=0.11h+8.5 \delta_ \nu$) for the case of high mass loading is shown in Fig. \ref{fig:RDF_wall}. Comparing between Figs. \ref{fig:RDF_wall}(a) and (c), the computed RDF based on four-way coupling point-force method is weaker than that based on the volume-filtering method. The streamwise RDF decreases to unity (unity indicates a random distribution) at $r/h>2$ ($r^+>454$) whereas the spanwise RDF decreases to unity at $r/h \sim 0.2$ ($r^+ \sim 46$), which indicates that the anisotropy content of particle cluster behaviour in the streamwise and spanwise directions. 

Meanwhile, we also calculate the RDF of particles in a slab with thickness of $\Delta y^+=8$ and $30$. The spanwise and global RDFs are only slightly different of particles in slabs with different thickness (figure not shown), while the streamwise RDF decreases (respectively increases) with increasing (respectively decreasing) thickness of the slab as shown in Figs. \ref{fig:RDF_wall}(a) and (c) (as expected). The computed RDF is noticeably more correlated than the measured RDF in \cite{FongJFM2019}, especially near the wall, consistent with the box-counting analysis shown previously. 

Furthermore, Figs. \ref{fig:RDF_wall}(b) and (d) depict the two-dimensional correlation of particle pairs in both spanmwise and streamwise directions, based on cases $high\_4\_R$ and $high\_4\_C$, respectively. Clearly, the computed ADF by case $high\_4\_C$ is more correlated than case $high\_4\_R$, in both the streamwise and spanwise directions. The distance from the maximum ADF to minimum ADF in the spanwise direction is around $\Delta r = 0.4h$ ($\Delta r^+ = 91$), which is comparable to the well-known streak spacing ($\Delta z^+ \sim O(100)$) between low with high speed streaks in single-phase flow. Additionally, this strong spanwise correlation is consistent at wide angles, due to the well-organized alternating low and high speed streaks. \\

Away from the wall, in the central region of the channel, the turbulence tends to be more isotropic. Fig. \ref{fig:RDF_center} shows RDF and ADF of particles in a slab along center plane ($y=h-0.015h$ to $y=h+0.015h$) for the case with high mass loading. Similar to the near-wall particle RDFs, the simulated RDFs are clearly more correlated than the experimental RDFs of \cite{FongJFM2019} whereas the simulated RDFs based on both four-way coupling methods agree with each other. From the contours of the ADFs in Figs. \ref{fig:RDF_center}(b,d), there is significantly less anisotropy of particle clustering in the channel center than in the near-wall region.\\

\section{Conclusion}
\label{5 Conclusion}
In this study, we investigate a vertical turbulent channel flow laden with moderate inertia particles ($St^+=58.6$) in the regime of two-way coupling to four-way coupling, with gravity aligned in the streamwise direction. Computed results from two independent DNS codes (based on point-force and volume-filtering methods), are compared directly with experimental measurement. 

Initially, particles are distributed randomly across the turbulent channel and gradually achieve a steady state, characterized by the particle turbophoresis time scale. Based on numerical simulations, particle/particle and particle/wall collisions have a negligible effect on the particle development time scale at low mass loading $(\Phi_m=6 \times 10^{-3})$, while collisions tend to delay the particle development time scale at high mass loading $(\Phi_m=0.1)$. When there is a fixed particle layer attached to each wall, more particles drift towards the center region from the near wall region, and the computed particle concentration profile was similar to measurements obtained using walls on which particle depositions had occurred.

For low mass loading $(\Phi_m=6 \times 10^{-3})$, the two-way coupling approach and both four-way coupling codes give nearly the same statistical profiles (e.g., particle mean velocity profile, concentration profile, particle RMS fluctuation velocity, particle Reynolds shear stress, and particle velocity skewness). On the other hand, there are more particles near the wall in simulations than observed in the experiment, which perhaps indicates that turbophoresis is stronger in the computational models than in the experiment. In spite of this discrepancy, computed results agree well with the measurement away from the wall. However, with increased mass loading to $\Phi_m=0.1$, more particles drift towards the channel center, and the experimental measurement suggests a dramatic turbulence modification with mass loading which is not observed in numerical simulations (even laden with greater number of higher inertia particles).

Particle clustering behaviours are analysed by Vorono\"i diagram analysis, box counting, and radial distribution functions in the case of high mass loading. Particle preferential concentration is strengthened with use of four-way coupling. This result is opposite to that of previous investigations in \cite{li2001numerical} and \cite{nasr2009dns}, who do not consider streamwise gravitational settling. Additionally, the angular distribution function is calculated in order to gain insight into the anisotropy of particle clusters. In the near-wall region, particle preferential concentration is higher in the streamwise direction than in the spanwise direction, especially in numerical simulations. Furthermore, the correlation coefficient is higher with the use of the four-way coupling volume-filtering method than point-force method. In the near-wall region, the mean distance of particle pairs correlates to the well-known streak spacing between the alternating low and high speed streaks in the single-phase flow. However, in the center region, both the particle clusters and turbulence structures tend to be more isotropic than in the near-wall region seen in both experiments and numerical simulations.


Taken together, these observations point to unmet challenges in modeling the behavior of wall-bounded particle-laden flows as measured in the laboratory. The uncertainties associated to the experiments should not be underestimated: the test case in \citet{FongJFM2019} was designed to remove some of the confounding factors present in previous studies, specifically the static charge on the walls that leads to particle adhesion and unwanted roughness. However, it is possible that unquantified triboelectric effects were still present, affecting particle-wall and particle-particle interactions. The incomplete streamwise development and residual turbophoretic drift may also have impacted the statistics. The finite channel aspect ratio (8:1) would not be considered a major factor in a single-phase flow, but the exact impact on the particle-laden case has not been addressed. In general, even in a relatively simple setting, to exactly identify all the consequential factors and all the important physical processes remains a challenge.\\

On the other hand, point-particle DNS has well known limitations in accounting for two-way coupling between particles and fluid flows, which motivated a number of recent efforts \citep{horwitz2016accurate, ireland2017improving}. However, at the considered concentrations and particle Reynolds numbers, it is not obvious how the classical mechanisms by which particles would affect the turbulence (particle wakes, mass loading, enhanced dissipation, see \citet{balachandar2010turbulent}) may account for the observed discrepancies. The weaker tendency to cluster in the experiments is possibly an indication that the interactions between particles and turbulent structures are not well modeled, or that unaccounted physical effects (e.g. particle charge) play a significant role in the experiments. \\

We underline that the present study is among the very few to directly address quantitative differences in a one-to-one comparison between measurements and simulations at matching conditions. The results clearly show that further investigations are warranted to ultimately achieve reliable predictive models of particle-laden turbulent flows.

\section*{Acknowledgments}
\label{6 Acknowledgments}

The authors acknowledge grants G00003613-ArmyW911NF-17-0366 from the U.S. Army Research Office and N00014-16-1-2472 from the Office of Naval Research. Computational resources were provided by the High Performance Computing Modernization Program (HPCMP), and by the ND Center for Research Computing.

\section*{References}
\label{7 References}





\end{document}